\documentclass[structabstract]{aa}  

\usepackage{graphicx}
\usepackage{txfonts}
\usepackage{natbib}
\bibpunct{(}{)}{;}{a}{}{,}
\def\spose#1{\hbox to 0pt{#1\hss}}
\def\ltsimm{\mathrel{\spose{\lower 3pt\hbox{$\sim$}}
        \raise 2.0pt\hbox{$<$}}}
\def\gtsimm{\mathrel{\spose{\lower 3pt\hbox{$\sim$}}
        \raise 2.0pt\hbox{$>$}}}
\def\d{{\rm{d}}}
\def\Rsol{\hbox{${\rm\thinspace R_{\odot}}$}}
\def\Msol{\hbox{${\rm\thinspace M_{\odot}}$}}
\def\yr{{\rm\thinspace yr}}
\def\Msolpyr{\hbox{${\rm\Msol\yr^{-1}\,}$}}
\def\aap{{\rm A\&A}}
\def\apj{{\rm ApJ}}
\def\apjs{{\rm ApJS}}

\def\mnras{{\rm MNRAS}}

\def\araa{{\rm ARA\&A}}

\def\cpc{{\rm Comp.~Phys.~Comm.}}
\def\jcp{{\rm J.~Comput.~Phys}}

\begin{document}
   \title{The 2.35 year itch of Cygnus OB2 $\#$9}
   \titlerunning{The 2.35 year itch of Cygnus OB2 $\#$9. III.}

   \subtitle{III. X-ray and radio emission analysis based
     on three dimensional hydrodynamical modelling}

   \author{E.~R.~Parkin\inst{1,2}, J.~M.~Pittard\inst{1},
     Y. Naz\'{e}\inst{3}\thanks{Research associate FNRS}, and
     R.~Blomme\inst{4}}

   \institute{School of Physics and Astronomy, The University of
     Leeds, Woodhouse Lane, Leeds LS2 9JT, UK \and
     Research School of Astronomy and Astrophysics, Australian National University, Canberra, ACT
     2611, Australia \and
     D\'{e}partment AGO, Universit\'{e} de Li\`{e}ge, All\'{e}e
         du 6 Ao\^{u}t 17, B5c, B-4000 Li\`{e}ge, Belgium \and
         Royal Observatory of Belgium, Ringlaan 3, 1180 Brussels,
         Belgium \\
     \email{rossparkin@gmail.com} }

   \date{Received ----; accepted ----}

 
  \abstract
  {The wind-wind collision in a massive star binary system leads to
    the generation of high temperature shocks that emit at X-ray
    wavelengths and, if particle acceleration is effective, may
    exhibit non-thermal radio emission. Cyg OB2$\#$9 represents one of
    a small number of massive star binary systems in this
    class.}
  {X-ray and radio data recently acquired as part of a project to
    study Cyg OB2$\#$9 are used to constrain physical models of the
    binary system, providing in-depth knowledge about the wind-wind
    collision and the thermal, and non-thermal, emission arising from
    the shocks.}
      {We use a three-dimensional, adaptive mesh refinement simulation
        (including wind acceleration, radiative cooling, and the
        orbital motion of the stars) to model the gas dynamics of the
        wind-wind collision. The simulation output is used as the
        basis for radiative transfer calculations considering the
        thermal X-ray emission and the thermal/non-thermal radio
        emission.}
      {The flow dynamics in the simulation show that wind acceleration
        (between the stars) is inhibited at all orbital phases by the
        opposing star's radiation field, reducing pre-shock velocities
        below terminal velocities. To obtain good agreement with the
        X-ray observations, our initial mass-loss rate estimates
        require a down-shift by a factor of $\sim 7.7$ to
        $6.5\times10^{-7} \Msolpyr$ and $7.5\times10^{-7} \Msolpyr$
        for the primary and secondary star, respectively. Furthermore,
        the low gas densities and high shock velocities in Cyg OB2
        $\#$9 are suggestive of unequal electron and ion temperatures,
        and the X-ray analysis indicates that an (immediately
        post-shock) electron-ion temperature ratio of $\simeq 0.1$ is
        also required. The radio emission is dominated by
        (non-thermal) synchrotron emission. A parameter space
        exploration provides evidence against models assuming
        equipartition between magnetic and relativistic energy
        densities. However, fits of comparable quality can be attained
        with models having stark contrasts in the ratio of
        magnetic-to-relativistic energy densities. Both X-ray and
        radio lightcurves are largely insensitive to viewing
        angle. The variations in X-ray emission with orbital phase can
        be traced back to an inverse relation with binary separation
        and pre-shock velocity. The radio emission also scales with
        pre-shock velocity and binary separation, but to positive
        powers (i.e. not inversely). The radio models also reveal a
        subtle effect whereby inverse Compton cooling leads to an
        increase in emissivity as a result of the synchrotron
        characteristic frequency being significantly reduced. Finally,
        using the results of the radio analysis, we estimate the
        surface magnetic field strengths to be $\approx 0.3-52\;$G.}
      {}

   \maketitle
%

\section{Introduction}
\label{sec:intro}
   
Hot, luminous, massive stars drive powerful stellar winds. The
wind-wind collision arising in a binary system leads to the generation
of high temperature shocks as the wind kinetic energy is thermalized
\citep{Stevens:1992, Pittard:2009}. Such shocks exhibit a number of
characteristic signatures. Firstly, they are prolific emitters at
X-ray wavelengths \citep[e.g.][]{Pittard_Parkin:2010}. Second, in a
number of cases non-thermal emission is observed
\cite[e.g.][]{Dougherty:2000, vanLoo_phd:2005, DeBecker:2006,
  vanLoo:2006, DeBecker:2007, vanLoo:2008, Naze:2008, Montes:2009,
  Blomme:2010, Volpi:2011, Reitberger:2012, DeBecker:2013,
  Blomme:2014}, with high spatial resolution radio observations
placing the source of the emission coincident with the wind-wind
collision shocks \citep[e.g.][]{Dougherty:2005}. The implication of
this latter finding is that particles are being accelerated to
relativistic energies at the wind-wind collision shocks by some
mechanism, for example, diffusive shock acceleration. Reassuringly,
theoretical models of non-thermal emission from the wind-wind
collision region in massive binary star systems are able to reproduce
the observed radio flux \citep{Eichler_Usov:1993, Benaglia:2003,
  Dougherty:2003, Pittard_et_al:2006, Reimer:2006, Pittard:2006,
  Reimer:2009, Farnier:2011, Falceta-Goncalves:2012}. Indeed, the
advantage of (almost) orbital-phase-locked, cyclic variations in the
shock physics compared to the more typically studied example of
supernovae remnants (see, for example, the reviews by
\citeauthor{Malkov:2001}~\citeyear{Malkov:2001} and
\citeauthor{Blasi:2013}~\citeyear{Blasi:2013}) make massive star
binaries an excellent laboratory for studying wind acceleration,
interacting radiation fields, X-ray emission, and non-thermal radio
emission.

A prime example that fits into this class of X-ray and non-thermal
radio emitting massive binary is the O+O-star system Cyg OB2$\#$9
\citep{vanLoo:2008, Naze:2008, Naze:2010, Volpi:2011}. Our recent
campaign to study this object has so-far consisted of an analysis of
optical and X-ray data by \cite{Naze:2012}, and radio observations by
\cite{Blomme:2013}, which have refined the orbital solution and
provided initial estimates for the stellar parameters - see
Tables~\ref{tab:system_parameters} and
\ref{tab:stellar_parameters}. The variation of the X-ray flux with
orbital phase correlates with the inverse of binary separation
(i.e. $1/d_{\rm sep}$), consistent with an adiabatic wind-wind
collision. The slope of the radio spectrum between 1.6 and 5 GHz
indicates non-thermal emission which can be accounted for reasonably
well by a simple wind-wind collision model. The result of the
observational effort is a multi-epoch, multi-wavelength dataset that
can be used to place firm constraints on physical models of the
wind-wind collision dynamics, and X-ray and radio emission arising
from Cyg OB2$\#$9.

In this paper we model the wind-wind collision dynamics of Cyg
OB2$\#$9 using a three dimensional adaptive-mesh refinement (AMR)
simulation (including wind acceleration, radiative cooling, and
orbital motion), which then acts as the input to radiative transfer
calculations for the emergent X-ray and radio emission. The X-ray
emission analysis presents strong evidence for a low ratio of
electron-to-ion temperatures in the immediately post-shock gas, and a
substantial lowering of mass-loss rates compared to previous
estimates. Moreover, to reproduce the observed radio lightcurves
requires a high efficiency of particle acceleration, with some
indication of a shallower slope to the relativistic electron
distribution than the ``standard'' case (i.e. $p<2$). The remainder of
this paper is structured as follows: the hydrodynamical and radiative
transfer calculations are described in \S~\ref{sec:model}. A brief
overview of the Cyg OB2$\#$9 binary system is given in
\S~\ref{sec:cygob29}. The simulation dynamics are examined in
\S~\ref{sec:dynamics}, followed by an analysis of the X-ray emission
in \S~\ref{sec:xrays} and radio emission in \S~\ref{sec:radio}. A
discussion of our results, an estimation of stellar surface magnetic
field strengths, and possible future directions are given in
\S~\ref{sec:discussion}, after which we close with conclusions in
\S~\ref{sec:conclusions}.

   \begin{table}
     \begin{center}
       \caption[]{Adopted system parameters for Cyg
         OB2$\#9$. References are as follows: 1 = \cite{Blomme:2013},
         2 = \cite{Naze:2012}, 3 =
         \cite{vanLoo:2008}.} \label{tab:system_parameters}
       \begin{tabular}{lll}
         \hline
         Parameter & Value & Reference \\
         \hline
         Orbital period, $P_{\rm orb}$ (d) & 860 & 1 \\
         $a$ ($\Rsol$) & $1733$ & $-$ \\
         Eccentricity ($e$) & 0.71 & 1 \\
         Distance (kpc) & 1.45 & 2 \\
         ISM column ($10^{22}\rm{cm}^{-2}$) & 1.15 & 2 \\
         Inclination, $i$ & $60^{\circ}$ & 3 \\
         Line-of-sight, $\theta$ & $280^{\circ}$ & 2 \\
         \hline
       \end{tabular}
     \end{center}
   \end{table}

   \begin{table}
     \begin{center}
       \caption[]{Adopted stellar/wind parameters. $k$ and $\alpha$
         are the \cite{Castor:1975} line driving
         parameters. References are as follows: 1 = \cite{Naze:2012},
         2 = \cite{Blomme:2013}. (Note that we have taken the median
         value from the ranges noted by \cite{Blomme:2013}.)
       } \label{tab:stellar_parameters}
        \begin{tabular}{lllll}
         \hline
          & \multicolumn{2}{c}{Primary star} & \multicolumn{2}{c}{Secondary star} \\
         \hline
         Parameter & Value & Reference & Value & Reference \\
         \hline
         Spectral Type & O5-O5.5I & 1 & O3-O4III  & 1  \\
         $M$ (M$_{\odot}$) & 50 & 1 & 44 & 1 \\ 
         $R_{\ast}$ (R$_{\odot}$) & 20 & 2 & 16 & 2 \\
         $T_{\rm{eff}}$ (K)& 37800 & 2 & 42200 & 2 \\
         $\log (L_{\ast}/L_{\odot}) $  & 5.85 &  2 & 5.87 & 2 \\
         $k$ & 0.24 & $-$ & 0.21 & $-$ \\
         $\alpha$ & 0.56 & $-$ & 0.60 & $-$ \\
         $\dot{M}\;({\rm M_{\odot}~yr^{-1}})$ & $5\times 10^{-6}$ & 2 & $5.8\times 10^{-6}$ & 2 \\
         $v_{\infty}\;({\rm km~s^{-1}})$ & 2060 & 2 & 2400 & 2 \\
         \hline
       \end{tabular}
    \end{center}
   \end{table}


\section{The model}
\label{sec:model}

\subsection{Hydrodynamic modelling}
\label{subsec:hydromodel}

The wind-wind collision is modelled by numerically solving the
time-dependent equations of Eulerian hydrodynamics in a 3D Cartesian
coordinate system. The relevant equations for mass, momentum, and
energy conservation are:
\begin{eqnarray}
\frac{\partial\rho}{\partial t} + \nabla \cdot (\rho {\bf v}) &  =  & 0, \\
\frac{\partial\rho{\bf v}}{\partial t} + \nabla\cdot(\rho{\bf vv}) + \nabla P & = & \rho{\bf f},\\
\frac{\partial\rho E}{\partial t} + \nabla\cdot[(\rho E + P){\bf v}] & =& \left(\frac{\rho}{m_{\rm H}}\right)^{2}\Lambda(T) + \rho {\bf f}\cdot {\bf v}.
\end{eqnarray}

\noindent Here $E = U_{\rm th}/\rho + \frac{1}{2}|{\bf v}|^{2}$, is the total
gas energy, $U_{\rm th}$ is the internal energy density, ${\bf v}$ is
the gas velocity, $\rho$ is the mass density, $P$ is the pressure, $T$
is the temperature, and $m_{\rm H}$ is the mass of hydrogen. We use
the ideal gas equation of state, $P = (\Gamma - 1) U_{\rm th}$, where
the adiabatic index $\Gamma = 5/3$.

The radiative cooling term, $\Lambda(T)$, is calculated from the
\textsc{APEC} thermal plasma code \citep{Smith_APEC:2001} distributed
in \textsc{XSPEC} (v12.5.1). Solar abundances are assumed
for the stellar winds \citep{Anders:1989}. The temperature of the
un-shocked winds is assumed to be maintained at $\approx 10^{4}\;$K via
photoionization heating by the stars.

The body force per unit mass ${\bf f}$ acting on each hydrodynamic
cell is the vector summation of gravitational forces from each star,
and continuum and line driving forces from the stellar radiation
fields. The calculation of the line force has been described by
\cite{Pittard:2009} and \cite{Parkin:2011b}, and we refer the reader
to these works for further details \citep[see also - ][]{Cranmer:1995,
  Gayley:1997}. In brief, the numerical scheme incorporates the
\cite*{Castor:1975} formalism for line driving by evaluating the local
Sobolev optical depth $t= \sigma_{\rm e}u_{\rm th}\rho[\hat{\bf
  n}\cdot \nabla ({\bf \hat{n}\cdot v)}]^{-1}$ \citep{Sobolev:1960} and
then calculating the vector radiative force per unit mass
\begin{equation}
{\bf g}_{\rm rad} = \frac{\sigma_{e}^{1-\alpha}k}{c} \oint I(\hat{\bf
  n})\left(\frac{\hat{\bf n}\cdot \nabla (\hat{\bf n}\cdot {\bf v})}{\rho
    v_{\rm th}}\right)^{\alpha} \left[ \frac{(1 + \tau_{\rm
      max})^{1-\alpha}}{\tau_{\rm max}^{1-\alpha}}\right]\hat{\bf n}d{\bf \Omega},\label{eqn:grad}
\end{equation}
\noindent where $\alpha$ and $k$ are the standard \cite{Castor:1975}
parameters, $\sigma_{\rm e}$ is the specific electron opacity due to
Thomson scattering, and $v_{\rm th}$ is a fiducial thermal velocity
calculated for hydrogen. The terms in square brackets in
Eq~(\ref{eqn:grad}) are included to capture the flattening of the line
force for small $t$ \citep{Owocki:1988}, where $\tau_{\rm max}= t
\eta_{\rm max}$ and we evaluate $\eta_{\rm max}$ from the 1D code used
to compute the initial wind profiles. A gaussian integration is
performed to correct the line force for the finite size of the stellar
disk \citep[][]{Castor:1974, Pauldrach:1986}. The line driving is set
to zero in cells with temperatures above $10^{6}\;$K, since this
plasma is mostly ionised.

The influence of X-ray ionisation on the line force parameters is not
included in the simulation \citep[i.e. ``self regulated shocks''
-][]{Parkin_Sim:2013}. We estimate the maximum reduction in pre-shock
wind speeds due to the feedback from post-shock X-rays ionising the
wind acceleration region \citep[using the model of][]{Parkin_Sim:2013}
to be a relatively minor alteration, on the order of $3\%$.

\subsection{The hydrodynamic code}
\label{subsec:hydrocode}

The hydrodynamic equations are solved using v3.1.1 of the
\textsc{FLASH} code \citep{Fryxell:2000, Dubey:2009}. The code uses
the piecewise-parabolic method of \cite{Colella:1984} to solve the
hydrodynamic equations and operates with a block-structured AMR grid
\citep[e.g.][]{Berger:1989} using the \textsc{PARAMESH} package
\citep{MacNeice:2000} under the message-passing interface (MPI)
architecture. A two-shock Riemann solver is used to compute the
inter-cell fluxes \citep{Colella:1984, Fryxell:2000}. The simulation
domain extends from $x = y = z =\pm2.06\times10^{14}\;$cm with outflow
boundary conditions used on all boundaries. The side-length of the
simulation domain equates to 2 (6) binary separations when the stars
reside at apastron (periastron). The grid is initialised with $x
\times y \times z = 16 \times 16 \times 16$ cubic blocks each
containing $8^{3}$ cells. We allow for 6 nested levels of refinement,
which results in an effective resolution on the finest grid level of
$8096^3\;$cells. The refinement of the grid depends on a
second-derivative error check \citep{Fryxell:2000} on $\rho$ and the
requirement of an effective number of cells between the stars
($\simeq100$ in this case) to accurately describe the WCR dynamics
\citep{Parkin:2011b}. Refinement and de-refinement are performed on
primitive variables\footnote{ Supplementary tests were performed to
  examine the influence of refinement on primitive or conserved
  variables on the conservation of mass, momentum and energy, as well
  as on the wind-wind collision dynamics and resulting X-ray
  emission. A negligible difference was observed.}  - we have found
this to reduce spurious pressure fluctuations at coarse-fine mesh
boundaries which arise in the high Mach number winds. Customised units
have been implemented into the \textsc{FLASH} code for radiative
driving, gravity, orbital motion, and radiative cooling for
optically-thin plasma \citep[using the method described
in][]{Strickland:1995}. An advected scalar is included to allow the
stellar winds to be differentiated. The stellar winds are
reinitialised within a radius of $\sim 1.15\;{\rm R_{\ast}}$ around
the stars after every time step. The orbital motion of the stars is
calculated in the centre of mass frame. When the stars are at apastron
the primary and secondary stars are situated on the positive and
negative $x-$axis, respectively. The motion of the stars proceeds in
an anti-clockwise direction.

\subsection{Radiative transfer calculations}
\label{subsec:RT}

\subsubsection{Adaptive image ray-tracing}

Synthetic X-ray/radio spectra and lightcurves are produced by solving
the equation of radiative transfer through the simulation domain using
adaptive image ray-tracing (AIR). For details of the {\sc AIR} code
the reader is referred to \cite{Parkin:2011}. In brief, an initially
low resolution image equivalent to that of the base hydrodynamic grid
(i.e. $128\times128\;$pixels) is constructed. The image is then
scanned using a second-derivate truncation error check on the
intrinsic flux to identify sufficiently prominent pixels for
refinement. In the present work we use $\xi_{\rm crit} = 0.6$, where
$\xi_{\rm crit}$ is the critical truncation error above which pixels
are marked for refinement - see \cite{Parkin:2011}. The process of
ray-tracing and refining pixels is repeated until features of interest
in the image have been captured to an effective resolution equivalent
to that of the shocked gas in the simulation. ( Note that this is a
separate post-simulation step, thus the radiative transfer
calculations do not influence the simulation dynamics.)

\subsubsection{Streamline integration}
\label{subsec:streamline}

To account for non-equilibrium electron and ion temperatures in the
X-ray emission calculations, and inverse Compton and Coulomb cooling
of relativistic electrons in the non-thermal radio calculations, we
require knowledge of the evolution of certain quantities. This
information can be retrieved by integrating along flow streamlines
using the simulation data. The first step in this process is to locate
the shocks, which we begin by locating the shocked gas in the
wind-wind collision region. Namely, those cells highlighted as
over-dense (i.e. post shock-jump gas) by a comparison of their density
to that computed from a wind following a $\beta$-velocity law with
$\beta=0.8$. It is then straightforward to locate the shocks that
align the shocked gas region(s). Local values of velocities are then
used to determine the flow direction and take incremental steps along
the streamline. Once this process has been completed for all cells
aligning the shocks, we recursively diffuse the scalars until all
cells containing shocked gas have been populated with a value for the
respective scalars.

Note that our approach of integrating along flow trajectories in
individual time snapshots from the simulations is an integration along
streamlines and not along path lines, i.e. we use the instantaneous
velocity field rather than the flow history. This approximation is
justified by the fact that the wind velocities are considerably larger
than the orbital velocity of the stars, hence changes to the shape of
the wind-wind collision region happen much faster than changes in the
position of the stars. We note that performing streamline integration
using an instantaneous velocity field will encounter difficulties in
un-steady, disconnected regions of post-shock gas. Fortunately, in the
present study this only becomes an issue for gas far downstream from
the apex of the wind-wind collision at phases close to periastron, for
which the contribution to both X-ray and radio emission is negligible.

\subsubsection{X-ray emission}
\label{subsubsec:xray_emission}

To calculate the X-ray emission from the simulation we use
emissivities for optically thin gas in collisional ionisation
equilibrium obtained from look-up tables calculated from the
\textsc{MEKAL} plasma code \citep{Mewe:1995, Kaastra:1992} containing
200 logarithmically spaced energy bins in the range 0.1-10 keV, and
101 logarithmically spaced temperature bins in the range
$10^{4}-10^{9}\;$K. When calculating the emergent flux we use energy
dependent opacities calculated with version $c08.00$ of Cloudy
\citep[][see also
\citeauthor{Ferland:1998}~\citeyear{Ferland:1998}]{Ferland:2000}. The
advected scalar is used to separate the X-ray emission contributions
from each wind\footnote{The post-shock gas temperatures and densities
  of the primary's and secondary's wind do not differ considerably for
  the wind-wind collision of Cyg OB2$\#$9. As such, diffusive
  numerical heating will have a negligible affect on results
  \citep{Parkin_Pittard:2010}.}.

Non-equilibrium electron and ion temperatures are expected for Cyg
OB2$\#$9 due to the low density and slow rate of electron heating via
Coulomb collisions in the post-shock gas - these points are discussed
further in \S~\ref{sec:cygob29}. We do not, however, include
non-equilibrium electron and ion temperatures, or ionisation effects,
in the hydrodynamic simulation as it would place too large a
computational strain on performing a comprehensive parameter space
study. Instead, we use the hydrodynamic simulation as a representative
realisation of the gas dynamics, and non-equilibrium electron and ion
temperatures are accounted for during the post-processing radiative
transfer calculations. The approach we use follows
\cite{Borkowski:1994} and \cite{Zhekov:2000}, whereby for a specific
X-ray calculation we define a value for the immediately post shock
ratio of the electron-to-ion temperature,
\begin{equation}
  \tau_{\rm e-i} = \frac{T_{\rm e}}{T} \bigg|_0,
\end{equation}
where $T_{\rm e}$ is the electron temperature. As the gas flows
through the shocks the electrons will be heated due to Coulomb
collisions. To account for this we integrate the following equation
along streamlines in the post shock gas,
\begin{equation}
  \frac{\d \tau_{\rm e-i}}{\d t} = (1 + x_{\rm e}) \frac{1 - \tau_{\rm e-i}}{t_{\rm
      eq}(T) \tau_{\rm e-i}^{3/2}}, \label{eqn:tet}
\end{equation}
where $x_{\rm e}$ is the relative electron fraction with respect to
the nucleons, and the timescale for temperature equilibration between
electrons and ions is taken to be \citep{Spitzer:1962},
\begin{equation}
  t_{\rm eq} = 252 \frac{\mu_{\rm i}}{Z^2} \frac{1}{\ln \Lambda_{\rm Coul}}
  \frac{T^{3/2}}{n_{\rm i}}, \label{eqn:teq}
\end{equation}
where $\Lambda_{\rm Coul}$ is the Coulomb logarithm and $n_{\rm i}$ is
the ion number density. Once values of $\tau_{\rm e-i}$ are known throughout
the post shock winds the ray-tracing calculations can be performed
with $T_{\rm e}$ used when computing the X-ray emission.

\subsubsection{Radio emission}

The radio emission calculations closely follow the methods outlined by
\cite{Dougherty:2003}, \cite{Pittard_et_al:2006}, and
\cite{Pittard:2006}. This model includes thermal and non-thermal
emission and absorption processes (free-free and synchrotron
emission/absorption and the Razin effect). The synchrotron emission
and absorption is taken to be isotropic, which follows from the
assumption that the magnetic field is tangled in the post-shock
gas. The effects on the relativistic electron distribution from
inverse Compton and Coulomb cooling, and the subsequent influence on
the emissivity, are also included. Note that we assume that the winds
are smooth, i.e., not clumpy - a discussion of the implications of
clumpy winds for the models results is given in
\S~\ref{subsec:clumpy_winds}. The main development in the present work
has been modifying the radio calculations to deal with the
three-dimensional adaptive-meshes used in the hydrodynamic
simulation. In essence, this involved taking the pre-existing {\sc
  AIR} code that was originally written for X-ray emission
calculations \citep{Parkin:2011} and incorporating the radio emission
calculations developed by \cite{Pittard_et_al:2006} - see also
\cite{Dougherty:2003}.

In the tangled magnetic field limit the synchrotron power at
frequency, $\nu$ for a single relativistic electron, is
\citep{Rybicki:1979, Pittard_et_al:2006},
\begin{equation}
  P_{\rm syn}(\nu) = \frac{\sqrt{3} q^3 B v}{m_e c^3} F(\nu/\nu_c), \label{eqn:synchrotron}
\end{equation}
where $q$ is the electron charge, $c$ is the speed of light, $B$ is
the magnetic field, and $F(\nu/\nu_c)$ is a dimensionless function
describing the shape of the synchrotron spectrum - tabulated values of
$F(\nu/\nu_c)$ can be found in \cite{Ginzburg:1965}. The characteristic
frequency for synchrotron emission (in the isotropic limit) is given by,
\begin{equation}
  \nu_c = \frac{3 \gamma^2 q B}{4 \pi m_e c} = \frac{3}{2} \gamma^2
  \nu_{b}, \label{eqn:nu_char}
\end{equation}
where $\nu_{b} = \omega_{b}/2\pi$, and $\omega_{b} = q B/m_e c$ is the
cyclotron frequency of the electron. The total synchrotron power is
then calculated by multiplying the emission per electron by the number
density of relativistic electrons. Note that the power per electron
depends on the magnetic field strength and on the Lorentz factor
through $F(\nu/\nu_c)$. 

The basic procedure is to use the hydrodynamic simulation to determine
appropriate values for thermodynamic variables such as density,
temperature, and pressure. This provides the necessary ingredients for
computing the thermal (free-free) radio emission/absorption along rays
traced through the simulation domain. To compute the non-thermal radio
emission/absorption we require three parameters to be defined. The
first is the power-law slope of the relativistic electron
distribution, $p$. In our model we suppose that the relativistic
electron distribution arises from first-order Fermi acceleration at
the wind-wind collision shocks. Test particle theory
\citep{Axford:1977, Bell:1978, Blandford:1978} tells us that electrons
accelerated at strong shocks (i.e. a compression ratio of 4) by this
process will result in a relativistic electron distribution with
$p=2$. The integrated number density of relativistic electrons is then
$\int_{\gamma_1}^{\gamma_2} n(\gamma) \d \gamma = C \gamma^{-p} \d
\gamma $, where $\gamma$ is the Lorentz factor, and $C$ sets the
normalisation of the distribution. For the intrinsic (i.e. un-cooled)
relativistic electron distribution we take $\gamma_{1}=1$ and
$\gamma_2=10^5$ (consistent with the maximum Lorentz factor attained
from the competition between acceleration by diffusive shock
acceleration and inverse Compton cooling). In reality, the maximum
Lorentz factor will vary along the shocks. However, for the cooled
electron distribution, most emission comes from electrons with
lower-than-maximum Lorentz factors, therefore we may reasonably adopt
a single representative value in the calculations.

The second parameter which we must define is the ratio of the
relativistic-to-thermal particle energy densities, $\zeta_{\rm rel} =
U_{\rm rel}/U_{\rm th}$, where $U_{\rm rel} = \int n(\gamma) \gamma
mc^2 \d \gamma$, $n(\gamma)$ is the number density of relativistic
electrons with Lorentz factor $\gamma$, and $m$ is the particle
mass. The value of $\zeta_{\rm rel}$ sets the normalisation of the
synchrotron emission and absorption \citep[see equations 6, 8, 9 and
11 in][]{Dougherty:2003}. For example, for a relativistic electron
distribution with $p=2$, the normalisation is $C = U_{\rm rel}/m_e c^2
\ln \gamma_{\rm max}$, where $\gamma_{\rm max}$ is the maximum
attainable Lorentz factor \citep[which is set by acceleration and
cooling processes, e.g. inverse Compton, see -][for further
discussion]{Dougherty:2003, Pittard_et_al:2006}. The third parameter
required in the radio calculations is the ratio of the
magnetic-to-thermal energy density, $\zeta_{\rm B} = (B^2/8\pi)/U_{\rm
  th}$. This parameter is necessary as the hydrodynamic simulation of
Cyg OB2$\#$9 does not explicitly provide magnetic field
information. The parameter $\zeta_{\rm B}$ directly influences the
normalisation of the radio spectrum ($P_{\rm syn} \propto B$), the
characteristic turn-over frequency of the spectrum ($\nu_{\rm c}
\propto B$), and the cut-off frequency caused by the Razin effect
($\nu_{\rm R} \propto B^{-1}$).

Our model calculations also account for the alteration to intrinsic
radio emission from inverse Compton cooling, Coulomb cooling, and the
Razin effect.  Due to the relatively close proximity of the shocks to
the intense UV radiation field of the stars, inverse Compton cooling
can significantly alter the relativistic electron distribution. The
rate of energy loss of a relativistic electron exposed to an isotropic
distribution of photons by inverse Compton scattering, in the Thomson
limit, is \citep{Rybicki:1979},
\begin{equation}
  P_{\rm compt} = \frac{4}{3} \sigma_{\rm T} c \gamma^2 \left(
    \frac{v}{c}\right)^{2} U_{\rm ph},
\end{equation}
where $\sigma_{\rm T}$ is the Thomson cross-section (assuming elastic
scattering and neglecting quantum effects on the cross-section) and
$U_{\rm ph}$ is the energy density of photons. Defining the cumulative
radiative flux from both stars, $f_{\rm stars} = L_1/r_1^2 + L_2/r_2^2
\;({\rm erg~s^{-1}~cm^{-2}})$, where $L_{i}$ and $r_{i}$ are the
luminosity and distance of a point from star $i$, respectively. The
time rate of change of the Lorentz factor due to inverse Compton
cooling, $\d \gamma / \d t|_{\rm IC}$ can then be evaluated from,
\begin{equation}
  \frac{\d \gamma}{\d t}\bigg|_{\rm IC} = \frac{\sigma_{\rm T} \gamma^{2}}{3
    \pi m_{\rm e} c^2}f = 8.61 \times 10^{-20} \gamma^2 f_{\rm stars} {\rm
    s}^{-1}. \label{eqn:IC}
\end{equation}
Note the dependence $\d \gamma / \d t|_{\rm IC} \propto \gamma^{2}$;
inverse Compton cooling is more significant for the highest energy
electrons - see \cite[][]{Pittard_et_al:2006} for further
discussion. The impact of inverse Compton cooling is assessed in our
model by integrating Eq~\ref{eqn:IC} along streamlines in the
post-shock region (see \S~\ref{subsec:streamline}).

In addition to inverse Compton cooling, as relativistic electrons flow
downstream away from the shocks they may lose energy through Coulomb
collisions with thermal ions. The rate of energy loss for electrons
from Coulomb cooling is (see
\citeauthor{Pittard_et_al:2006}~\citeyear{Pittard_et_al:2006} and
references there-in),
\begin{equation}
  \frac{\d \gamma}{\d t} \bigg|_{\rm Coul} = 3\times10^{-14}\frac{n_{\rm
      i}\ln \Lambda}{\sqrt{\gamma^{2} -1}} {\rm s^{-1}}. \label{eqn:CC}
\end{equation}
In contrast to inverse Compton cooling, Coulomb cooling is more
significant for lower energy electrons as $\d \gamma/ \d t |_{\rm
  Coul} \propto (\gamma^{2} -1)^{-1/2}$. The impact of Coulomb cooling
on the relativistic electron distribution may be readily assessed by
integrating Eq~\ref{eqn:CC} along streamlines (as outlined in
\S~\ref{subsec:streamline}).

The intrinsic radio emission may also be reduced by the Razin
effect. In essence, the Razin effect relates to a suppression of the
``beaming'' of synchrotron emission, which is caused by a change in
the refractive index of the plasma surrounding the emitting
region. This effect is incorporated in our models by replacing the
intrinsic Lorentz factor for relativistic electrons (after inverse
Compton and Coulomb cooling) with the modified value,
\begin{equation}
  \gamma' = \frac{\gamma}{\sqrt{1 + \gamma^2 \nu_{0}^2/\nu^2}},
\end{equation}
where $\nu$ is the emission frequency, $\nu_{0} = \sqrt{q^2 n_{\rm
    e}/\pi m_{\rm e}}$ is the plasma frequency, and $q$ is the
electron charge.

For further information regarding the calculations for free-free
emission/absorption and synchrotron emission/absorption we refer the
reader to \cite{Dougherty:2003}, \cite{Pittard_et_al:2006}, and
\cite{Pittard:2006}.

\section{Cyg OB2$\#$9}
\label{sec:cygob29}

Before proceeding to the results of the hydrodynamic simulation and
emission calculations, in this section we will recap the results of
previous studies and highlight important physics for models of Cyg
OB2$\#$9. The adopted parameters of the system and stars are noted in
Tables~\ref{tab:system_parameters} and \ref{tab:stellar_parameters},
respectively. The large variation in observed radio flux across the
orbit \citep{vanLoo:2008, Blomme:2013} is characteristic of a highly
eccentric binary system. Indeed, the orbital solution derived by
\cite{Naze:2010} returned $e\sim 0.744$. This has since been revised
to $e \simeq 0.711$, with an orbital period of $\simeq 860\;$days,
using the recent optical and radio analysis by \cite{Naze:2012} and
\cite{Blomme:2013}. The model constructed by \cite{Naze:2012} (based
on estimates for the stellar parameters) provides valuable physical
insight into the system, particularly the anticipated significance of
wind acceleration on the importance of radiative cooling. These models
show that at orbital phases close to periastron the combined effect of
the shocks entering the wind acceleration regions and the opposing
stars radiation field reducing the net acceleration cause the
post-shock gas to be on the border-line between radiative and
quasi-adiabatic \citep[see also][]{Parkin_Gosset:2011}. This has a
knock-on effect on the plasma temperature, reducing it below that
estimated for models where stellar winds are instantaneously
accelerated at the stellar surfaces. Based on the above, we may expect
large changes in the character of the shocks between apastron and
periastron, which will be conveyed in the observed X-ray and radio
emission.

The wind densities for Cyg OB2$\#$9 are sufficiently low as to afford
the question of whether the shocks are collisionless or
collisional. Importantly, there is the possibility that electron
temperatures may be significantly lower than ion temperatures in the
immediately post-shock region, increasing as gas flows downstream and
Coulomb interactions bring about electron-ion temperature
equilibration. Analyses of X-ray emission from supernova remnants has
revealed a trend of decreasing electron-to-ion temperature ratio,
$T_{\rm e}/T$, with increasing shock velocity
\citep{Rakowski:2005}. For the shock velocities in Cyg OB2$\#$9,
typical values of $T_{\rm e}/T \simeq 0.1-0.2$ are anticipated. One
may estimate the importance of non-equilibrium electron and ion
temperatures\footnote{Alternatively, one may use the scalar
  constructed by \cite{Zhekov:2007} to estimate the importance of
  non-equilibrium ionisation.} in Cyg OB2$\#$9 by considering the
timescale for electron-ion temperature equilibration, $t_{\rm eq}$
(see Eq~\ref{eqn:teq}), and the length scale for Coulomb collisional
dissipation, $l_{\rm Coul} \simeq 7\times10^{18} v_{8}^4/n_{\rm i}$
\citep{Pollock:2005}. In Table~\ref{tab:NEI} we list values of $t_{\rm
  eq}$ and $l_{\rm Coul}$ for Cyg OB2$\#$9, where simulation data has
been used to estimate $n_{\rm i}$ and $T$. For our initial mass-loss
rate estimates of $\dot{M}_1=5\times10^{-6}\;{\rm M_{\odot}~yr^{-1}}$
and $\dot{M}_2=5.8\times10^{-6}\;{\rm M_{\odot}~yr^{-1}}$, the
computed value of $t_{\rm eq}=0.12\;P_{\rm orb}$, when contrasted with
the flow time for the post-shock winds of $t_{\rm flow}\sim d_{\rm
  sep}/v_{\infty} \simeq 0.01\;P_{\rm orb}$, suggests that an
electron-ion temperature ratio less than unity may occur for
Cyg~OB2$\#$9. Furthermore, if lower mass-loss rates are considered,
the equilibration timescale and length scale for Coulomb collisional
dissipation increase. For example, adopting
$\dot{M}_1=6.5\times10^{-7}\;{\rm M_{\odot}~yr^{-1}}$ and
$\dot{M}_2=7.5\times10^{-7}\;{\rm M_{\odot}~yr^{-1}}$, one finds
$t_{\rm eq} = 0.92 P_{\rm orb}$. Hence, there is a larger likelihood
for non-equilibrium electron and ion temperatures in models with lower
(than our initial estimates) mass-loss rates.

   \begin{table}
     \begin{center}
       \caption[]{Importance of non-equilibrium ionisation for Cyg
         OB2$\#$9. The binary separation at apastron, $d_{\rm sep}
         (\phi= 0.5)=2965\Rsol$, and at periastron $d_{\rm sep} (\phi=
         1)=501\Rsol$. System and stellar parameters are noted in
         Tables~\ref{tab:system_parameters} and
         \ref{tab:stellar_parameters}, respectively.} \label{tab:NEI}
        \begin{tabular}{lllll}
         \hline
         $\dot{M}_1$ & $\dot{M}_2$ & $t_{\rm eq}/P_{\rm orb}$ &
         $l_{\rm Coul}/ d_{\rm sep} $ & $l_{\rm Coul}/
         d_{\rm sep} $ \\
         $({\rm M_{\odot}~yr^{-1}})$ & $({\rm M_{\odot}~yr^{-1}})$ & & $(\phi= 0.5)$
         & $(\phi= 1)$ \\
        \hline
        $5\times10^{-6}$ & $5.8\times10^{-6}$ & 0.12 & 0.005 & 0.029 \\
        $6.5\times10^{-7}$ & $7.5\times10^{-7}$ & 0.92 & 0.038 & 0.223 \\
        $5\times10^{-7}$ & $5.8\times10^{-7}$ & 1.20 & 0.050 & 0.290 \\
        \hline
       \end{tabular}
    \end{center}
   \end{table}

\begin{figure*}
  \begin{center}
    \begin{tabular}{cc}
     \resizebox{80mm}{!}{\includegraphics{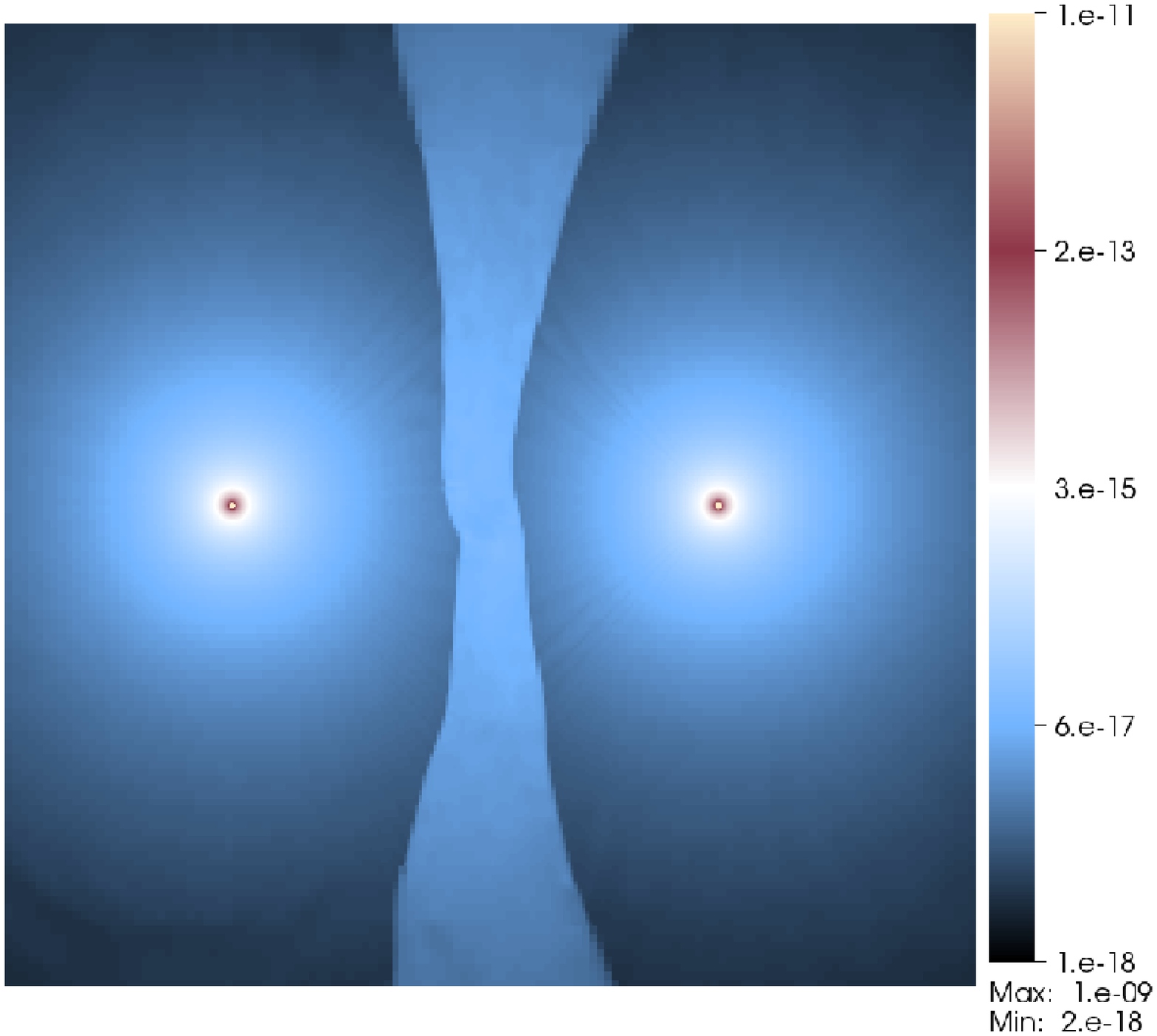}} &
     \resizebox{80mm}{!}{\includegraphics{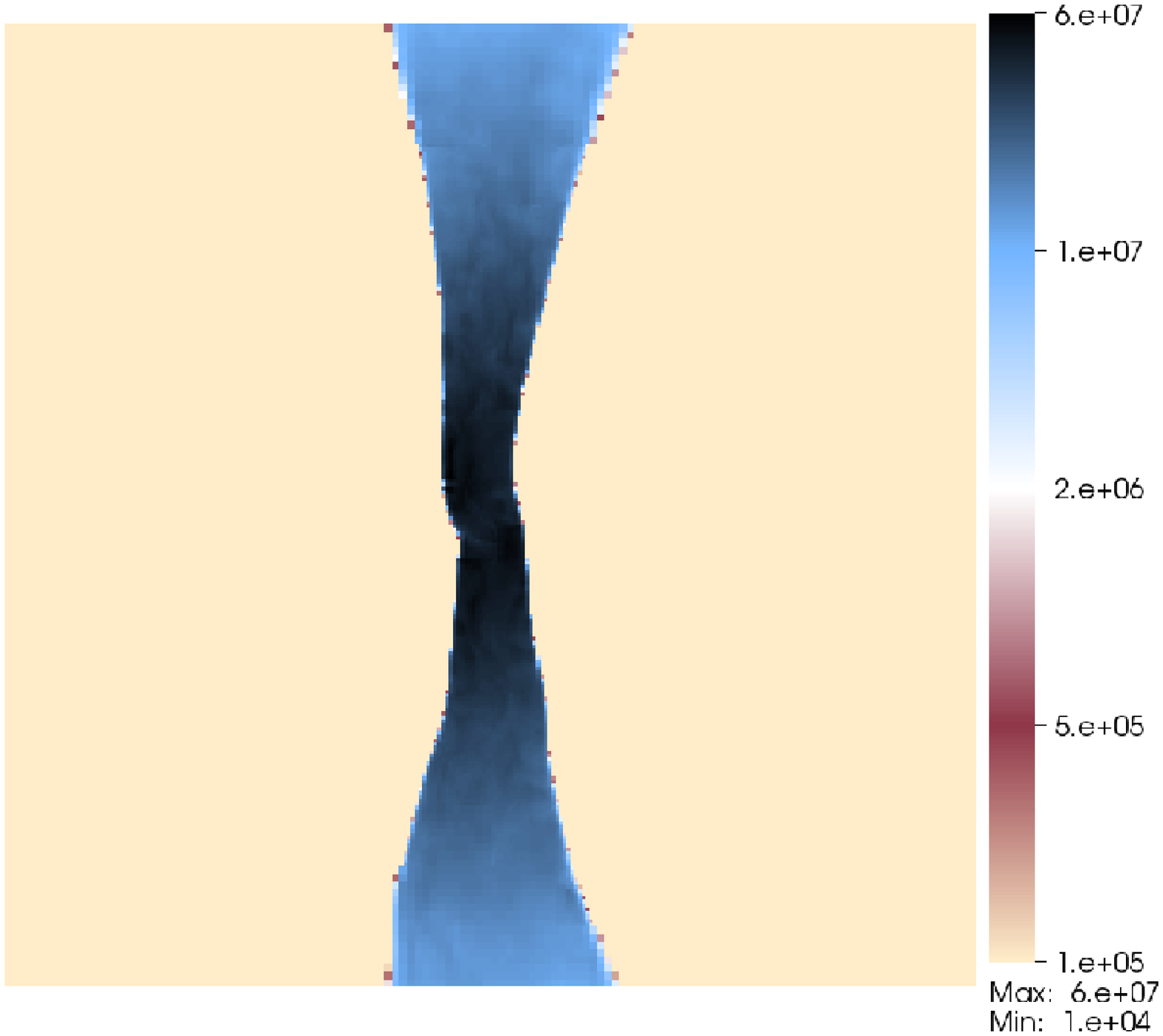}} \\
     \resizebox{80mm}{!}{\includegraphics{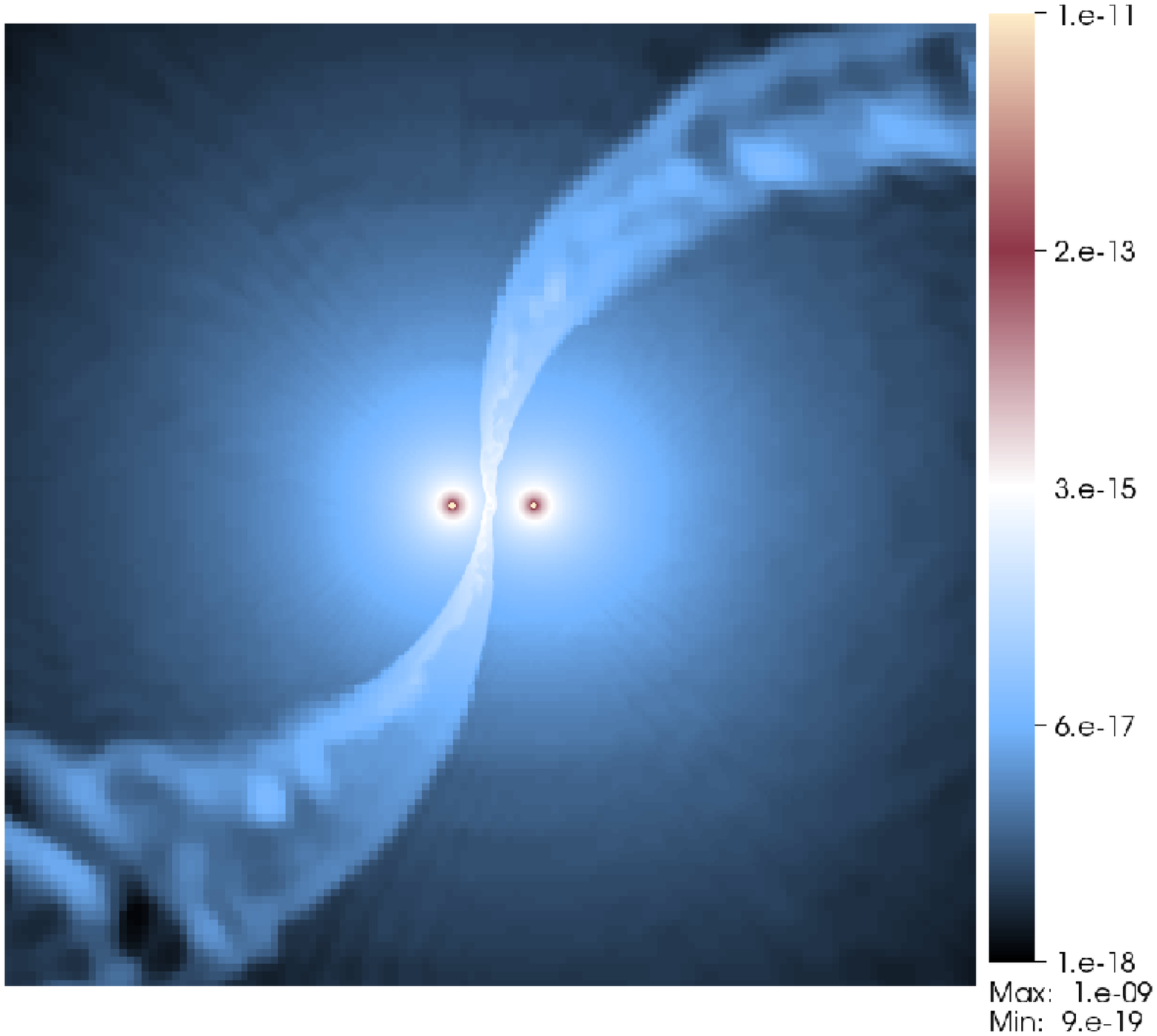}} &
     \resizebox{80mm}{!}{\includegraphics{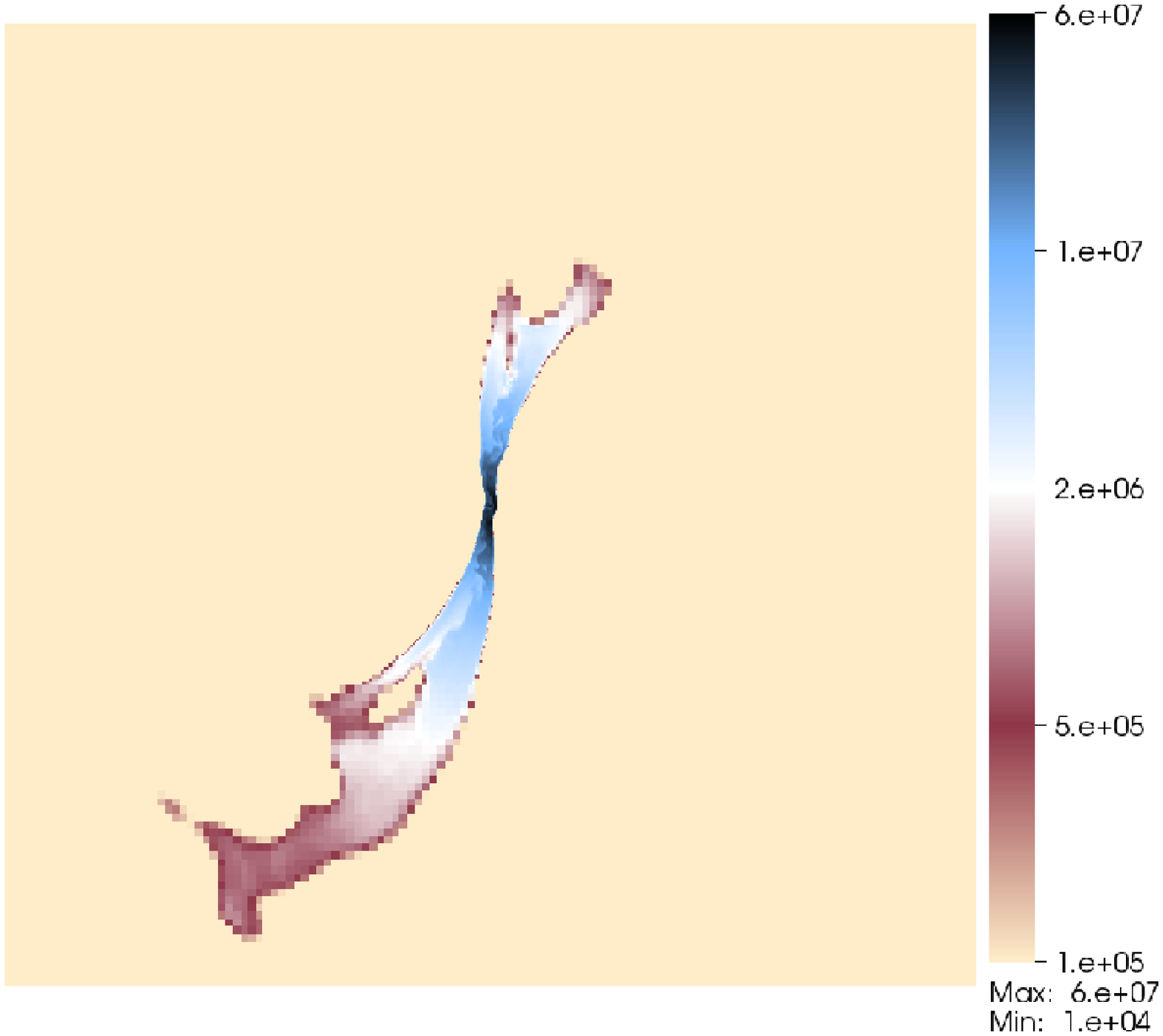}} \\
   \end{tabular}
   \caption{Simulation snapshots showing the orbital ($x-y$) plane at
     orbital phases, $\phi=0.5$ (top) and 1.0 (bottom). The left and
     right columns show density (in units of g cm$^{-3}$) and
     temperature (in K), respectively. All plots show the full $x-y$
     extent of the domain $=\pm2.06\times10^{14}\;$cm $=2963\Rsol$ -
     for comparison, the primary and secondary star have a radius of
     20 and $16\Rsol$, respectively (see \S~\ref{subsec:hydrocode} for
     further details). At apastron (top row) the primary star is to
     the right, and vice-versa at periastron (bottom row).}
   \label{fig:snapshots}
 \end{center}
\end{figure*}

\begin{figure*}
  \begin{center}
    \begin{tabular}{ccc}
    \resizebox{170mm}{!}{\includegraphics{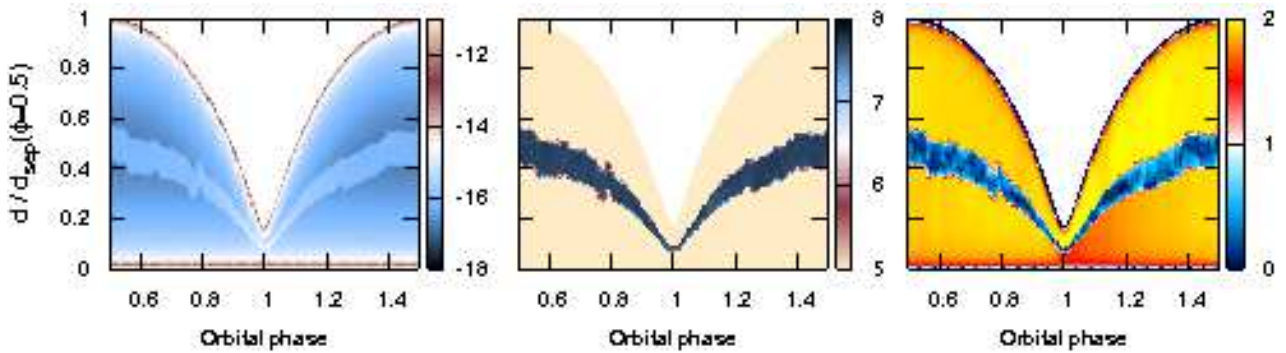}} \\
   \end{tabular}
   \caption{Data extracted from the line-of-centres as a function of
     orbital phase. The primary star is situated at $d_{\rm sep}=0$,
     where $d_{\rm sep}$ is the binary separation, and the
     phase-dependent position of the secondary star coincides with the
     ``V''-shape in the plots. Shown are: logarithm of density in g
     cm$^{-3}$ (left), logarithm of temperature in K (middle), and
     line-of-centres velocity in units of $10^{8}\;$cm s$^{-1}$
     (right).}
   \label{fig:loc}
 \end{center}
\end{figure*}

\section{Hydrodynamics of Cyg~OB2$\#$9}
\label{sec:dynamics}

\subsection{Qualitative features of the simulation}

Snapshots of the density and temperature at apastron ($\phi=0.5$) and
periastron ($\phi=1$) from the hydrodynamic simulation of Cyg OB2$\#$9
are shown in Fig.~\ref{fig:snapshots}. A number of interesting
qualitative features are highlighted. At orbital phases close to
apastron the motion of the stars is relatively slow in comparison to
the wind velocities and as such the influence of orbital motion on the
colliding-winds region (CWR) is minor. The collision of the stellar
winds leads to the production of strong shocks between the stars with
temperatures reaching $\simeq 6\times10^{7}\;$K. The relatively high
temperatures of post-shock gas as it flows off the grid shows that
radiative cooling is not important at apastron.

As the stars approach periastron the binary separation contracts and
the relative speed of the stars increases. The influence of orbital
motion, namely the curvature to the shocks resulting from the Coriolis
force \citep[see also, for example,][]{Parkin:2008, Parkin:2009,
  Pittard:2009, vanMarle:2011, Lamberts:2012, Madura:2013}, is
apparent in the arms of the colliding winds region. However, the shape
of the shocks within a binary separation of the apex remains largely
the same as at apastron, indicating that the wind momentum ratio does
not change considerably through the orbit. At both apastron and
periastron the CWR is roughly equidistant between the stars
(Fig.~\ref{fig:snapshots}), consistent with a stand-off distance
estimate (assuming terminal velocity winds) of, $r = d_{\rm sep}
\eta^{1/2}/(1+\eta^{1/2}) = 0.46 d_{\rm sep}$, with $\eta =
\dot{M}_{1} v_{\infty 1}/\dot{M}_2 v_{\infty 2} = 0.74$
(Table~\ref{tab:stellar_parameters}).

An indication of the change in separation of the stars through the
orbit can be gained from Fig.~\ref{fig:loc} where we show the density,
temperature, and wind velocity along the line-of-centres between the
stars as a function of orbital phase. The position of the shocks
between the stars can be seen as the V-shaped feature in the figure;
the post-shock gas density increases at smaller binary separations,
when the shocks reside closer to the stars.  Although the post-shock
gas temperature (on the line-of-centres) does not change considerably
through the orbit (middle panel of Fig.~\ref{fig:loc}), it is clear
from the lower right panel of Fig.~\ref{fig:snapshots} that the
downstream post-shock gas cools effectively at phases close to
periastron. In the downstream, curved arms of the CWR, where the
temperature $\ltsimm 10^{5}\;$K, the gas is unstable, exhibiting a
similar flow corrugation to that observed in simulations of
WR\thinspace22 by \cite{Parkin_Gosset:2011}.

\subsection{Wind velocities}

The simulation includes wind acceleration and the additional effect of
the opposing stars radiation field in accelerating/decelerating the
wind. As such, we may expect two effects to arise: radiative
inhibition \citep{Stevens:1994}, and radiative braking
\citep{Gayley:1997}. Radiative inhibition refers to the reduction in
the net acceleration of the winds between the stars as a result of the
opposing stars radiation field. Radiative braking, on the other hand,
refers to the sharp deceleration of a wind as it approaches the
photosphere of the star (with the weaker wind). This effect is
expected to be important for systems with disparate wind strengths,
whereby the CWR occurs close to one of the stars. Given this,
radiative braking is not as important as radiative inhibition for Cyg
OB2 $\#$9 because the shocks are roughly equidistant from the stars.

The rightmost panel in Fig.~\ref{fig:loc} shows the wind velocity
along the line-of-centres between the stars as a function of orbital
phase. The sharp contrast in colour close to the stars shows the rapid
acceleration of each stellar wind - the primary star resides on the
x-axis and the secondary star is coincident with the V-shape. Also
apparent in the figure is an asymmetry in the wind velocities either
side of periastron. For example, the velocities in the pre-shock
primary star's wind are noticeably higher at $\phi=0.9$ than at
$\phi=1.1$. In contrast, the wind velocities appear higher after
periastron passage in the secondary star's wind. To examine these
features in more detail, in Fig.~\ref{fig:vpre_chi} we show the
pre-shock wind velocity as a function of orbital phase. For
comparison, estimates from 1D wind acceleration calculations
\citep[e.g.][]{Stevens:1994} are shown - note that these estimates
differ from those presented by \cite{Naze:2012} because of the
different stellar parameters adopted in this work. Immediately
noticeable is that the wind velocity along the line-of-centres does
not reach the terminal velocity (Table~\ref{tab:stellar_parameters})
at any phase through the orbit. Moreover, the pre-shock velocities of
both winds do not reach the values estimated from 1D model
calculations of static stars that include competing radiation
fields. At periastron the primary's wind exhibits a pronounced dip in
excess of that anticipated from the 1D estimate, very reminiscent of
that observed in simulations of $\eta\thinspace$Car by
\cite{Parkin:2011b} \citep[see also][]{Madura:2013}. A considerable
deviation from the 1D estimate is observed for the secondary star's
wind, with the pre-shock velocity showing a small increase rather than
a decrease at periastron. Whether this is the result of a reduction in
the inhibiting influence of the primary star, or an enhancement to the
wind acceleration is not clear. Returning to Fig.~\ref{fig:loc}, there
is an indication from the plot that the variation in pre-shock
velocity at orbital phases near to periastron arises due to a
variation in wind velocities closer to the shocks than to the
stars. Put another way, the sharp gradient in the wind profile close
to the star (corresponding to the inner wind acceleration region) does
not change considerably throughout the orbit. The implication of this
finding would be that the variation in pre-shock velocities seen in
Fig.~\ref{fig:vpre_chi} is the result of orbital motion affecting the
winds at a distance greater than a few stellar radii from the stars. A
detailed analysis of the contributions from the relative motion of the
stars and orbital phase dependent velocity projection effects
\citep[e.g.][]{Parkin:2011b} is beyond the scope of the current
work. 

\subsection{Importance of cooling}

The snapshots of density and temperature in Figs.~\ref{fig:snapshots}
and \ref{fig:loc} indicate that at phases close to periastron the
post-shock gas in the CWR is smooth and remains at a high temperature
($T\gtsimm 10^{6}\;$K) within a distance from the shock apex of at
least twice the binary separation. The significance of this is that
gas does not cool effectively in the region where the majority of the
observed X-ray and (non-thermal) radio emission is emitted. As such,
this region can be considered to be quasi-adiabatic throughout the
orbit. This contrasts with the downstream arms of the CWR,
particularly at periastron, where cooling is clearly effective.

\begin{figure}
  \begin{center}
    \begin{tabular}{c}
    \resizebox{80mm}{!}{\includegraphics{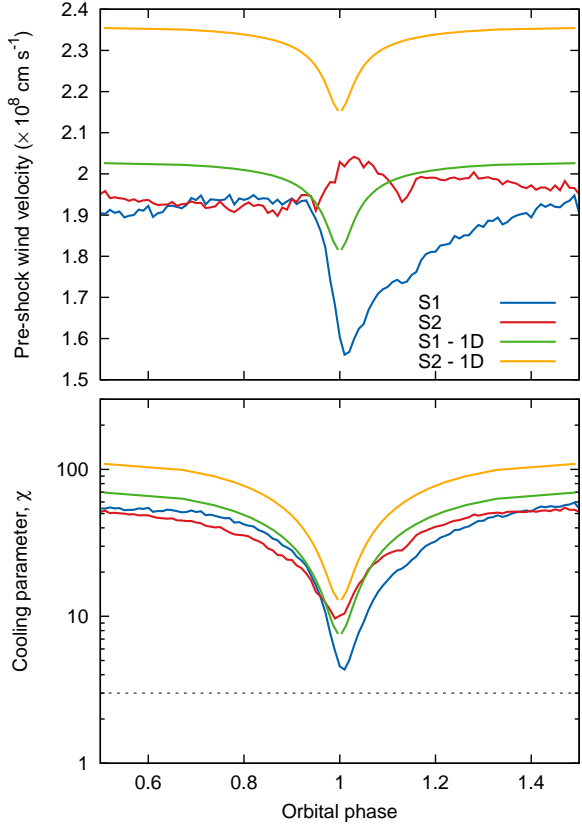}} \\
  \end{tabular}
  \caption{Pre-shock velocities (upper panel) and post-shock cooling
    parameters (lower panel) derived from the three-dimensional
    hydrodynamic simulation. For comparison, curves are shown for
    one-dimensional static-star model estimates (which include wind
    acceleration and the radiation fields of both stars). The dashed
    line in the lower panel corresponds to $\chi = 3$; when $\chi
    \ltsimm 3$ gas cools effectively within the post-shock region
    where emission predominantly originates from.}
   \label{fig:vpre_chi}
 \end{center}
\end{figure}

To place the assertions in the previous paragraph on a more
quantitative footing, we use the simulation data to compute the
cooling parameter for the post-shock gas, $\chi$. Following
\cite{Stevens:1992}, we define $\chi$ to be the ratio of the cooling
time to the flow time:
\begin{equation}
  \chi = \frac{t_{\rm cool}}{t_{\rm flow}} = \frac{v_{8}^4 d_{12}}{\dot{M}_{-7}},
\end{equation}
where $v_{8}$ is the pre-shock wind velocity (in units of $10^{8}\;$cm
s$^{-1}$), $d_{12}$ is the binary separation (in units of
$10^{12}\;$cm), and $\dot{M}_{-7}$ is the wind mass-loss rate (in
units of $10^{-7}\Msolpyr$). For reference, $\chi \gg 1$ denotes
adiabatic gas, $\chi \ll 1$ corresponds to rapidly cooling gas, and
$\chi \simeq 3$ means that the post-shock gas will be quasi-adiabatic
close to the CWR apex but will cool roughly three binary separations
downstream. The lower panel of Fig.~\ref{fig:vpre_chi} shows values of
$\chi$ as a function of orbital phase. Throughout the orbit $\chi > 4$
which means that the shock apex region remains quasi-adiabatic. Values
of $\chi$ for both winds show similar evolution, and the deviation
from the 1D model estimates can be attributed to the comparatively
lower pre-shock wind velocities and asymmetry about periastron passage
(see upper panel of Fig.~\ref{fig:vpre_chi}).

Noting the results of the X-ray emission analysis in the following
section, a key finding is that achieving acceptable fits to the column
density requires the wind mass-loss rates to be reduced by a factor of
$\simeq7-8$. This has the consequence of reducing the importance of
cooling in the winds because of the corresponding reduction in the gas
density and the simple fact that $\chi \propto 1/\dot{M}_{-7}$. As
such, for the lowered mass-loss rates that we consider later, cooling
is less important in the region of the CWR close to the shock apex
(where the majority of X-ray and radio emission originates
from). Consequently, although an entirely consistent approach would
require a separate simulation to be computed for each choice of
mass-loss rates, a very good approximation can be made using the
single simulation that we have performed, but with gas densities
adjusted to account for different mass-loss rates. 

\section{X-ray emission}
\label{sec:xrays}

In this section we confront the X-ray observations of \cite{Naze:2012}
with emission estimates from radiative transfer calculations with the
{\sc AIR} code, using density and temperature values from the
hydrodynamic simulation as input. For a description of the methods
used see \S~\ref{sec:model}. We focus in this section on using the
results of the radiative transfer calculations to constrain the wind
mass-loss rates and the ratio of electron-to-ion temperature. This
analysis reveals that our initial mass-loss rate estimates require a
downward revision, and that values of $T_{\rm e}/T\ll 1$ in the post
shock gas are supported by the fits to observations. All X-ray
calculations adopt viewing angles of $i=60^{\circ}$ and
$\theta=280^{\circ}$, where $i$ is the inclination angle and $\theta$
is the angle subtended against the positive $x$-axis (in the orbital
plane). Our adopted values of $i$ and $\theta$ are taken from
\cite{vanLoo:2008} and \cite{Naze:2012}, respectively. Note that when
comparing models to observations, all fits are performed ``by eye''.

\begin{figure}
  \begin{center}
    \begin{tabular}{c}
     \resizebox{80mm}{!}{\includegraphics{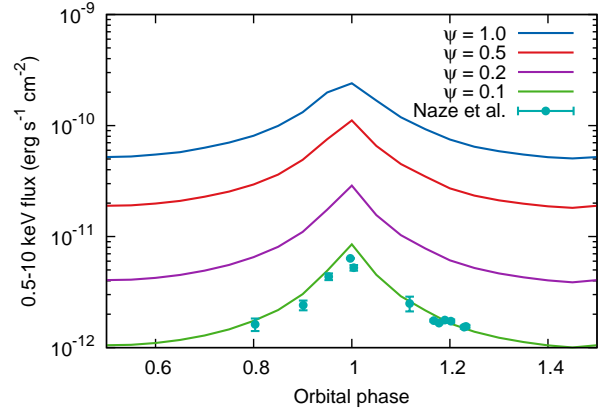}} \\
 \end{tabular}
 \caption{0.5-10 keV X-ray lightcurve compared against the observed
   fluxes derived by \cite{Naze:2012}. The different curves correspond
   to different reduction factors for the wind mass-loss rates, $\psi$
   (see Eq.~\ref{eqn:mdot}).}
   \label{fig:xray_lc1}
 \end{center}
\end{figure}

\begin{figure}
  \begin{center}
    \begin{tabular}{c}
    \resizebox{80mm}{!}{\includegraphics{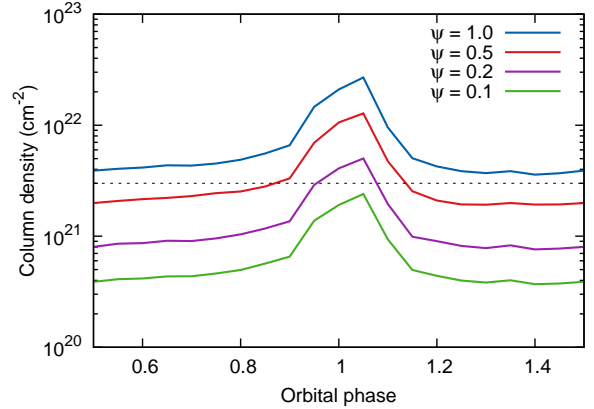}} \\
  \end{tabular}
  \caption{Column density as a function of orbital phase computed from
    the X-ray radiative transfer calculations. The different curves
    corresponds to different reduction factors for the wind mass-loss
    rates. The horizontal dashed line in the column density plot
    corresponds to the best-fit value of $3\times10^{21}~$cm$^{-2}$
    attained by \cite{Naze:2012} for phases close to periastron.}
   \label{fig:xray_EWC}
 \end{center}
\end{figure}

\subsection{Constraining wind mass-loss rates}
\label{subsec:mdots}

The first order of business is to assess how well the model using our
initial estimates for stellar and system parameters fares against the
observations. The blue curve in Fig.~\ref{fig:xray_lc1} shows the
0.5-10 keV flux computed from the model. The variation in the X-ray
luminosity as a function of orbital phase is consistent with the
scaling $L_{\rm X} \propto \rho^{2} \propto d_{\rm sep}^{-1} v_{\rm
  pre}^{-2}$ (where $v_{\rm pre}$ is the pre-shock wind velocity)
which follows from the assumption that $\dot{M}$, wind momentum ratio,
the post-shock gas temperature, and $\Lambda(T) \sim $ are roughly
constant. Note that $d_{\rm sep}$ changes more than $v_{\rm pre}$
between apastron and periastron, hence $L_{\rm X}$ scales more
strongly with $d_{\rm sep}$. The blue curve in
Fig. ~\ref{fig:xray_lc1} also quite obviously over-estimates the
observed 0.5-10 keV flux by a factor $\sim 50$. Similarly, the
standard model gives an emission weighted column density\footnote{The
  emission weighted column is defined as $\Sigma f(2-10\;{\rm keV})
  n_{\rm H} / \Sigma f(2-10\;{\rm keV})$, where $n_{\rm H}$ is the
  column density accrued along an individual sight line, $f(2-10\;{\rm
    keV})$ is the 2-10 keV X-ray flux, and the sums are over all sight
  lines in the radiative transfer calculations \citep{Parkin:2008}.}
that over-estimates the value of $3\times10^{21}\;$cm$^{-2}$ derived
from X-ray spectral fits for phases close to periastron by
\cite{Naze:2012} (illustrated by the dashed horizontal line in
Fig.~\ref{fig:xray_EWC}).

The over-estimates in column density suggest that our adopted wind
mass-loss rates are too high. Indeed, lowering the wind mass-loss
rates, which we simulate by reducing the gas density in the
calculations by a factor,
\begin{equation}
  \psi = \frac{\dot{M_{i}}}{\dot{M}_{i-0}}, \label{eqn:mdot}
\end{equation}
where $\dot{M}_{i-0}$ is our initial guess at the wind mass-loss rates
(Table~\ref{tab:stellar_parameters}), brings the model calculations
into much better agreement with our observational constraint of
$3\times10^{21}\;$cm$^{-2}$ (Fig.~\ref{fig:xray_EWC}). Furthermore,
reduced mass-loss rates also substantially improve the fit of the
0.5-10 keV X-ray flux (Fig.~\ref{fig:xray_lc1}). We find that a value
of $\psi \simeq 0.1$ is optimal for simultaneously fitting the column
density and X-ray fluxes.

It should be noted that since the post-shock winds are quasi-adiabatic
($\chi \gtsimm 4$ - see Fig.~\ref{fig:vpre_chi}) throughout the orbit,
the shocks will be relatively smooth and we cannot appeal to
thin-shell mixing, or oblique shock angles, caused by corrugated shock
interfaces as a mechanism to reduce the intrinsic X-ray luminosity as
these effects are typically associated with radiative shocks - see,
e.g., \cite{Parkin_Pittard:2010}, \cite{Owocki:2013}, and
\cite{Kee:2014}. Furthermore, although an under-estimate in the
observationally inferred column density\footnote{In this regard it is
  worth noting the results \cite{Pittard_Parkin:2010} and
  \cite{Parkin_Gosset:2011}, who showed that column density can be
  energy dependent and, moreover, the column density returned by
  standard fitting procedures may not be exactly correct.} would allow
larger mass-loss rates, the over-prediction of the X-ray luminosity
(which scales with the mass-loss rate) is unavoidable unless mass-loss
rates are reduced.

At first hand the suggested reduction in the mass-loss rates may seem
drastic. However, we note two factors in this regard. Firstly, there
is the uncertainty in wind mass-loss rates due to wind clumping
\citep[see, for example,][]{Puls:2008}. (Note, however, that clumpy
colliding winds in the adiabatic regime do not affect the X-ray flux
\citeauthor{Pittard:2007}~[\citeyear{Pittard:2007}].) Secondly,
considerable reductions to mass-loss rates were also implied by the
analysis of \cite{Parkin_Gosset:2011} for WR22, and for a sample of
small separation systems by \cite{Zhekov:2012}. From a more general
perspective, our initial mass-loss rate estimates are based on
theoretical calculations for single massive stars. Thus it is our goal
to see how well such estimates fare when applied to binary systems.

\subsection{Non-equilibrium electron and ion temperatures}

The stellar wind densities in Cyg OB2$\#$9 are sufficiently low that
there is potential for the shocks to be collisionless (see
\S~\ref{sec:cygob29}). The reduction to wind mass-loss rates inferred
in the previous section will increase this likelihood. Thus we may
reasonably expect that electrons will initially be at a lower
temperature than ions in the post shock gas.

\begin{figure}
  \begin{center}
    \begin{tabular}{c}
     \resizebox{80mm}{!}{\includegraphics{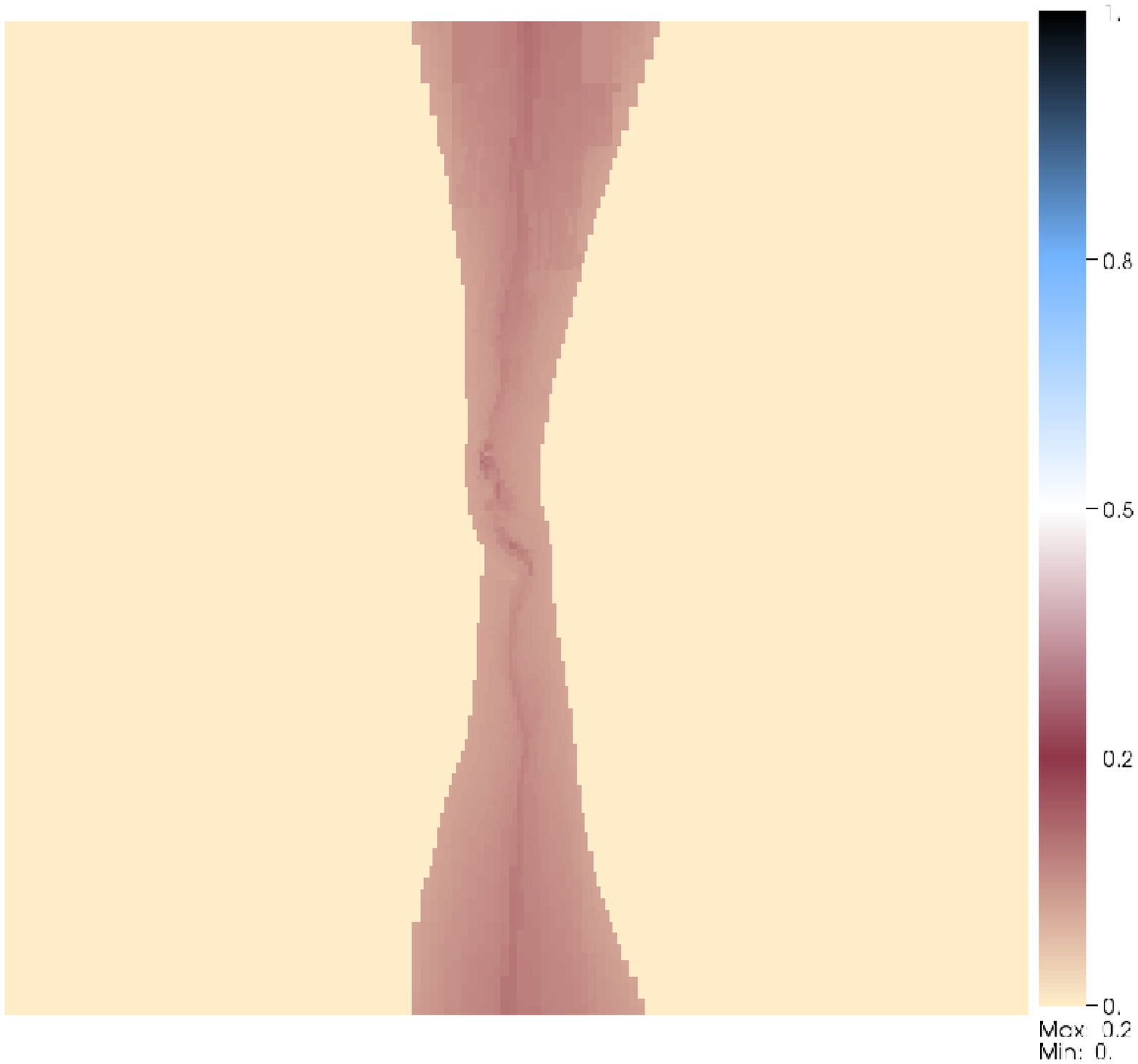}} \\
     \resizebox{80mm}{!}{\includegraphics{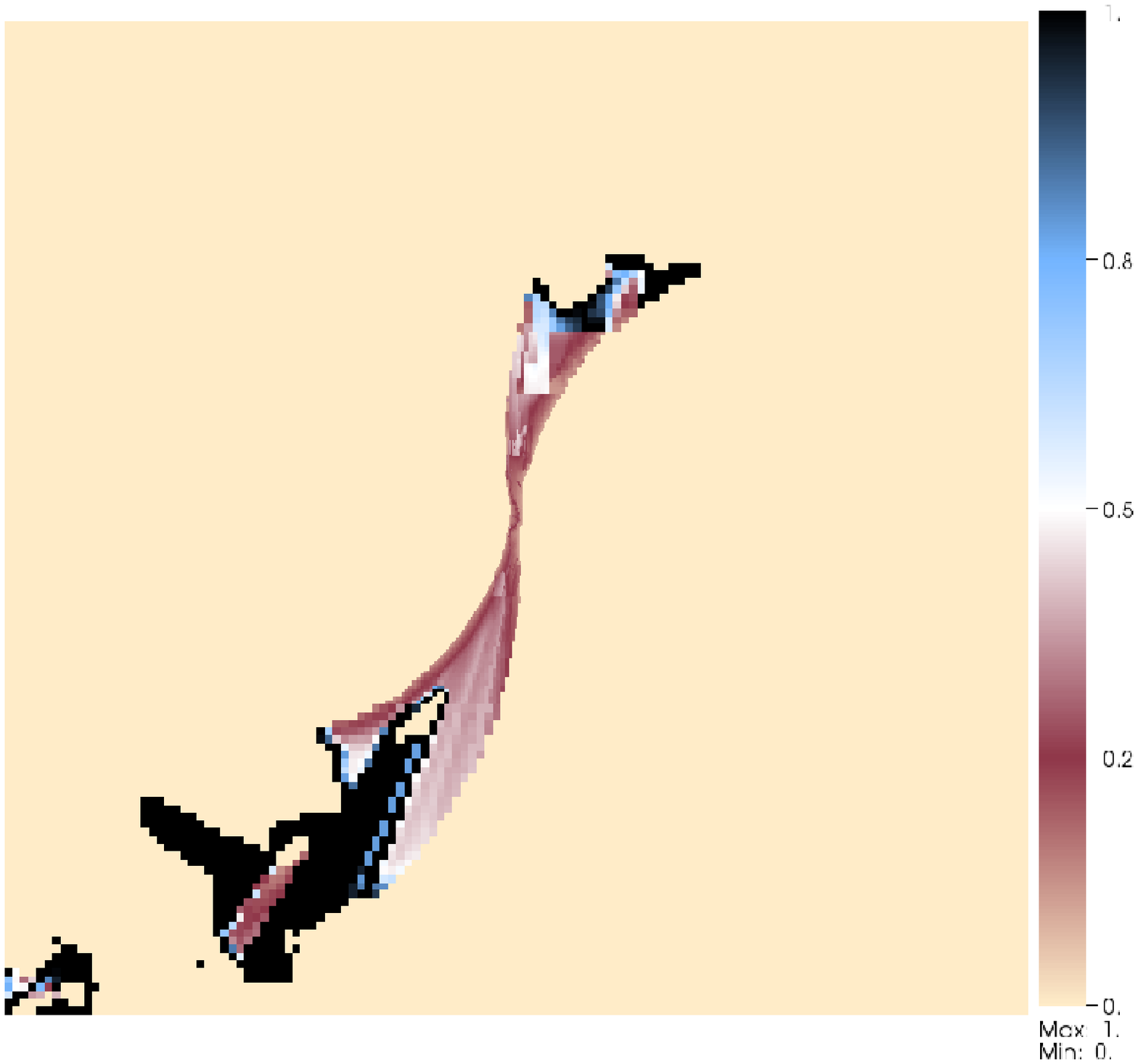}} \\
  \end{tabular}
  \caption{Spatial distribution of the ratio of electron-to-ion
    temperature, $T_{\rm e}/T$, from a model calculation with
    $\tau_{\rm e-i} (=T_{\rm e}/T|_0) = 0.1$ and $\psi = 0.1$. The
    orbital ($x-y$) plane is shown at phases, $\phi=0.5$ (top) and 1.0
    (bottom). All plots show the full $x-y$ extent of the domain (see
    \S~\ref{subsec:hydrocode}). Note that values of $T_{\rm e}/T$ in
    the un-shocked winds are plotted as zero to aid visibility but are
    taken to be unity in the X-ray calculations.}
   \label{fig:tet}
 \end{center}
\end{figure}

As a visual aid for the spatial dependence of electron temperatures in
our radiative transfer calculations, we show snapshots of $T_{\rm
  e}/T$ at apastron and periastron from a model with $\tau_{\rm
  e-i}=0.1$ (recall that $\tau_{\rm e-i}$ defines $T_{\rm e}/T$
immediately post shock - see \S~\ref{subsubsec:xray_emission}) and
$\psi=0.1$ in Fig.~\ref{fig:tet}. The figure shows that electrons are
heated as they flow away from the shocks. There is a noticeable rise
in electron temperatures at the contact discontinuity, due to the
increased time since this gas was shocked. At phases close to
apastron, post shock gas densities are at their lowest and $T_{\rm
  e}/T$ remains relatively low throughout the shocked winds region. In
contrast, at periastron (lower panel in Fig.~\ref{fig:tet}) Coulomb
collisions heat the electrons much quicker and at a few binary
separations downstream $T_{\rm e}/T$ has risen to close to a half,
with electron and ion temperatures reaching equilibration just prior
to where the shocks are disrupted by instabilities (see lower left
panel of Fig.~\ref{fig:snapshots}). Importantly, however, electron
temperatures are low in the region where X-ray emission predominantly
originates from (i.e., within a binary separation of the apex of the
wind-wind collision region).

\begin{figure}
  \begin{center}
    \begin{tabular}{c}
    \resizebox{80mm}{!}{\includegraphics{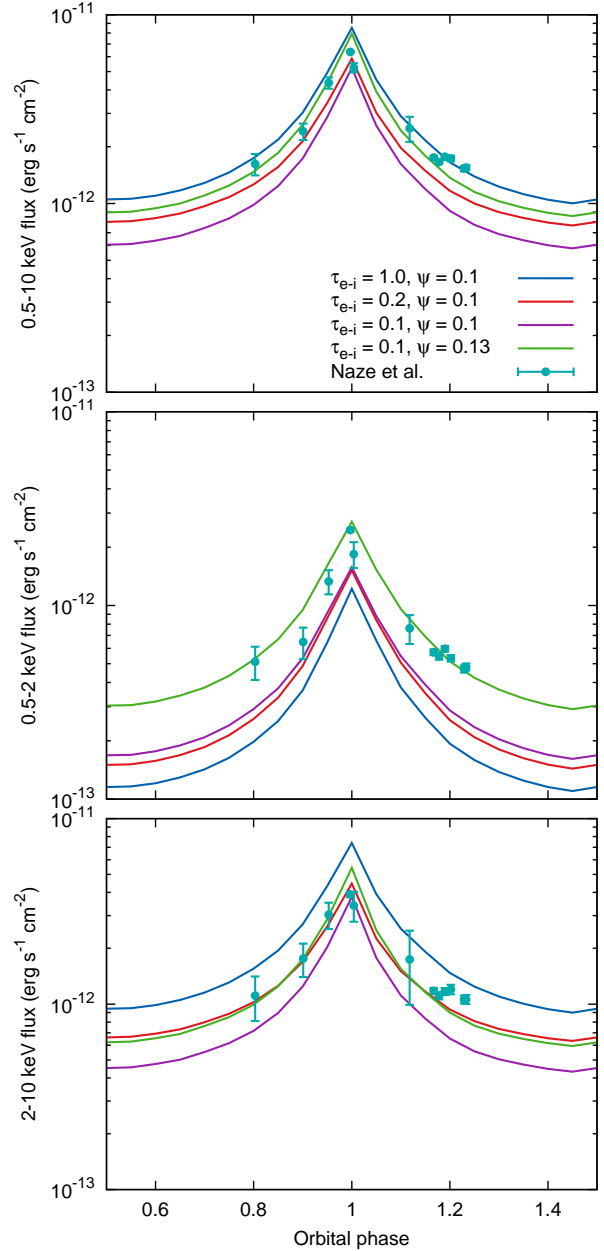}} \\
  \end{tabular}
  \caption{X-ray lightcurves examining the effect of reducing the
    value of $\tau_{\rm e-i} (=T_{\rm e}/T|_0)$ on the 0.5-10 (top),
    0.5-2 (middle), and 2-10 keV (bottom) fluxes. These calculations
    have reduced mass-loss rates, with the reduction factor ($\psi$)
    indicated in the plots. Observationally derived X-ray fluxes
    \citep{Naze:2012} are shown for comparison.}
   \label{fig:xray_lc2}
 \end{center}
\end{figure}

In Fig.~\ref{fig:xray_lc2} curves are plotted for a series of X-ray
models in which $\tau_{\rm e-i}$ has been varied. (Note that mass-loss
rates as implied from the analysis in \S~\ref{subsec:mdots} have been
used in these calculations.) Interestingly, without invoking
non-equilibrium electron and ion temperatures (i.e. when $\tau_{\rm
  e-i}=1$) the models underestimate the observed 0.5-2 keV X-rays
while overestimating the 2-10 keV flux. Reducing $\tau_{\rm e-i}$
remedies this deficiency, and a markedly better fit to the
observations is acquired with $\tau_{\rm e-i} \simeq 0.1$. In essence,
by reducing $\tau_{\rm e-i}$ the X-ray spectrum has been shifted to
lower energies, with an additional effect of lower energy X-rays being
more susceptible to absorption. This explains why the 0.5-10 keV flux
shows a steady decline with decreasing $\tau_{\rm e-i}$.

\begin{figure}
  \begin{center}
    \begin{tabular}{c}
    \resizebox{85mm}{!}{\includegraphics{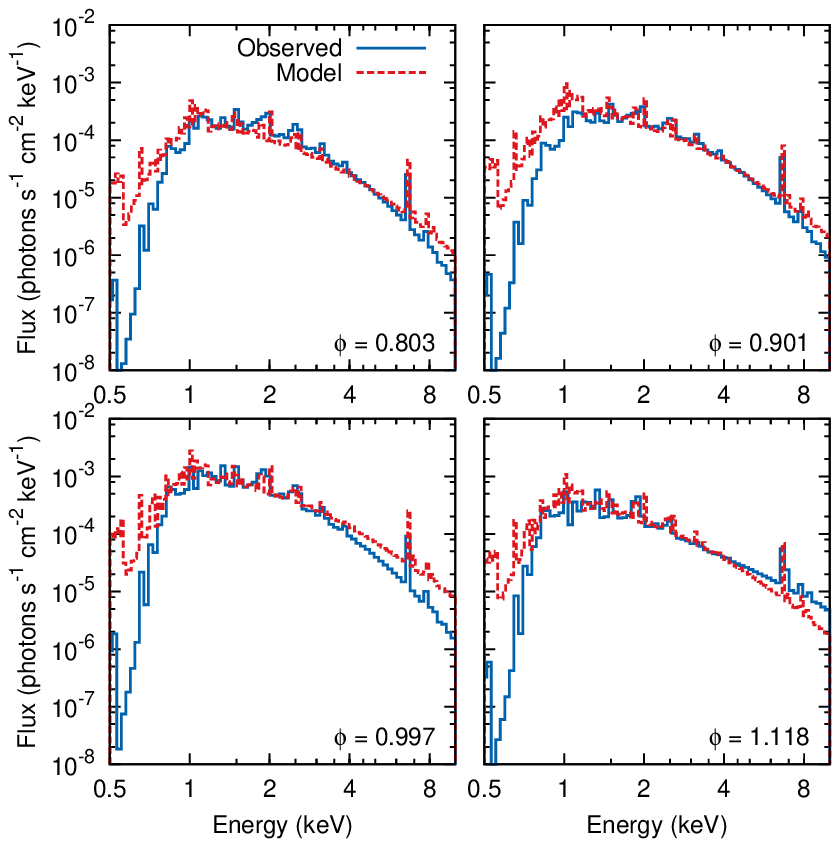}} \\
    \end{tabular}
   \caption{Comparison of the best-fit model X-ray spectra to
     observations. The model was calculated using reduced mass-loss
     rates ($\psi=0.13$; $\dot{M}_1 = 6.5\times10^{-7} \Msolpyr$ and
     $\dot{M}_2 = 7.5\times10^{-7} \Msolpyr$) and $\tau_{\rm
       e-i}=0.1$. The observed X-ray spectra at $\phi=0.803$, 0.901,
     and 1.118 were acquired with {\it SWIFT} and the $\phi=0.997$
     spectrum was acquired with {\it XMM-Newton}
     \citep{Naze:2012}. (The observed spectra correspond to the fits
     with ``one temperature fixed'' presented by
     \citeauthor{Naze:2012}~\citeyear{Naze:2012}.)  Orbital phases of
     the observations ($\phi$) are noted in the plots.}
   \label{fig:xray_spectra}
 \end{center}
\end{figure}

In Fig.~\ref{fig:xray_spectra} we compare the best-fit model spectra
against a sample of the observed {\it SWIFT} and {\it XMM-Newton}
spectra \citep{Naze:2012}. The fit of the model to the data is best in
the 1-4 keV energy range. At both lower and higher energies the model
and observations diverge slightly. The over-estimate in X-ray flux at
energies less than 1 keV could be remedied by an increase in
absorption, which could be achieved by a minor increase in the
mass-loss rates. The slightly higher model flux at $E \gtsimm 8$
compared to observations at $\phi=0.803$, 0.901, and 0.997 suggests
that the adopted wind velocities are a little too large. However, the
fact that the model under-predicts the high energy flux at
$\phi=1.118$ provides evidence to the contrary, and we are
apprehensive to adjust the adopted wind velocities. All in all, the
best-fit model provides a good match to the observed spectra, although
there is an indication that the agreement could be improved with some
minor adjustments to stellar, and perhaps also system, parameters. 

The above results indicate a significant importance for differing
electron and ion temperatures for Cyg OB2$\#$9. This value is in good
agreement with the X-ray analysis of WR140 by \cite{Zhekov:2000},
where $\tau_{\rm e-i}\sim 0.2$ was found to provide better fits of
model spectra to observations than $\tau_{\rm
  e-i}=1$. (\citeauthor{Pollock:2005}~[\citeyear{Pollock:2005}] found
further evidence of non-equilibrium electron and ion temperatures in
WR140 from the low velocity widths of less energetic ions, implying a
slow rate of post-shock electron heating.) In contrast,
\cite{Zhekov:2007} found that X-ray spectral fits for WR147 (which has
mass-loss rates closer to Cyg OB2$\#$9 than WR140, and a larger binary
separation) demanded $\tau_{\rm e-i} \simeq 1$ and rapid post-shock
temperature equilibration between electrons and ions. Attempting to
identify trends between systems where models including non-equilibrium
electron and ion temperatures have been considered is, therefore, not
a straightforward task. Both Cyg OB2$\#$9 and WR140 have large orbital
eccentricity, and similar binary separations, yet mass-loss rate
estimates for these two systems differ by more than an order of
magnitude. On the other hand, terminal wind velocities may provide a
divider as Cyg OB2$\#$9 and WR140 have $v_{\infty} >2000\;{\rm
  km~s^{-1}}$, whereas WR147 has $v_{\infty} \simeq2000\;{\rm
  km~s^{-1}}$. Considering the modest number of attempts to include
non-equilibrium electron and ion temperatures in models of colliding
wind binaries, and the broad range of post-shock electron-to-ion
temperature ratios inferred, it will important to build a more
complete sample to better inform our understanding of these systems in
future studies. It is, however, noteworthy that the initial post-shock
electron-ion temperature ratio of $\tau_{\rm e-i} \simeq 0.1$ for Cyg
OB2$\#$9 is in good agreement with values derived for X-ray emitting
supernovae blast waves \citep{Rakowski:2005}.

In summary, to fit the observed column density and X-ray flux we are
required to reduce our initial mass-loss rate estimates by a factor of
roughly 5-10 and invoke the presence of low temperature electrons in
the post shock gas. After a number of additional trials (not
documented herein), we arrived at $\psi =0.13$, giving best-fit values
of $\dot{M}_1 = 6.5\times10^{-7}\Msolpyr$ and $\dot{M}_2 =
7.5\times10^{-7}\;{\rm M_{\odot}~yr^{-1}}$, and an immediately
post-shock electron-ion temperature ratio, $\tau_{\rm e-i} \simeq
0.1$. The results of using these best-fit parameters is illustrated by
the green curve in Fig.~\ref{fig:xray_lc2}.

\section{Radio emission}
\label{sec:radio}

In this section we continue to confront our model of Cyg OB2$\#$9 with
observational constraints, now turning to the radio domain. Previous
models of Cyg OB2$\#$9 have provided tantalising evidence for
non-thermal radio emission from a wind-wind collision
\citep{vanLoo:2008, Blomme:2013}. In the following we proceed using
the mass-loss rates determined from \S~\ref{sec:xrays} and focus on
the following goals: establishing whether non-thermal radio emission
is required to match the observed radio flux; estimating the
relativistic electron and magnetic energy densities required in such a
case; and, investigating whether a non-standard power-law slope to the
relativistic electron distribution is supported (e.g. $p<2$). As with
the X-ray emission models in the previous section, fits of the radio
models to the observational data are ``by eye''.

\subsection{Inverse Compton cooling}
\label{subsec:IC}

Inverse Compton cooling has a significant impact on the distribution
of relativistic electrons in the post-shock gas, and thus affects the
emergent synchrotron emission. In this section we highlight the
spatial dependence of inverse Compton cooling in the post-shock gas
and the resulting radio emission. To set the scene for the parameter
space exploration that will follow, the example calculations are
performed using two contrasting sets of parameters: $\zeta_{\rm rel}
\gg \zeta_{\rm B}$ and $\zeta_{\rm rel} \ll \zeta_{\rm B}$.

\begin{figure}
  \begin{center}
    \begin{tabular}{c}
     \resizebox{80mm}{!}{\includegraphics{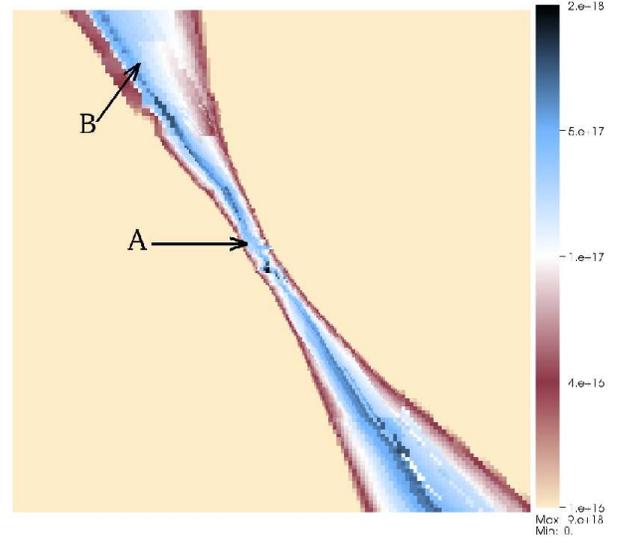}} \\
 \end{tabular}
 \caption{Spatial distribution of the inverse Compton cooling scalar
   at orbital phase, $\phi=0.8$ from a model calculation with
   $\zeta_{\rm rel} = 0.15$, $\zeta_{\rm B} = 5\times10^{-5}$, and
   $p=1.6$. The full $x-y$ extent of the domain is shown (see
   \S~\ref{subsec:hydrocode}). Points ``A'' and ``B'' are reference
   points for comparing the influence of inverse Compton cooling on
   the emergent radio emission (see \S~\ref{sec:radio}). Note that the
   scalar is zero in the un-shocked winds.}
   \label{fig:ics}
 \end{center}
\end{figure}

Fig.~\ref{fig:ics} shows the spatial distribution of $1/\gamma^2 \d
\gamma/\d t$, which is used to quantify the importance of inverse
Compton cooling in our calculations - see Eq~\ref{eqn:IC}. In essence
this scalar tracks the accumulated stellar flux incident upon parcels
of gas as they are advected through the CWR. Values are low
immediately post shock, but grow as parcels of gas are advected
through the post shock region and become increasingly illuminated by
the stellar radiation fields \citep{Pittard_et_al:2006}. A ``knot'' is
apparent at the stagnation point of the shocks (close to the centre of
the image in Fig.~\ref{fig:ics}). This feature indicates how flow
which is slowed in the post shock region is subject to a larger dose
of stellar radiation, thus limiting the maximum Lorentz factor of the
relativistic electrons to a greater degree. At a number of points
along the contact discontinuity which separate the shocks, one can see
similar patches. From studying movies of the simulation, these patches
are related to instabilities in the post shock gas.

\begin{figure}
  \begin{center}
    \begin{tabular}{c}
      \resizebox{85mm}{!}{\includegraphics{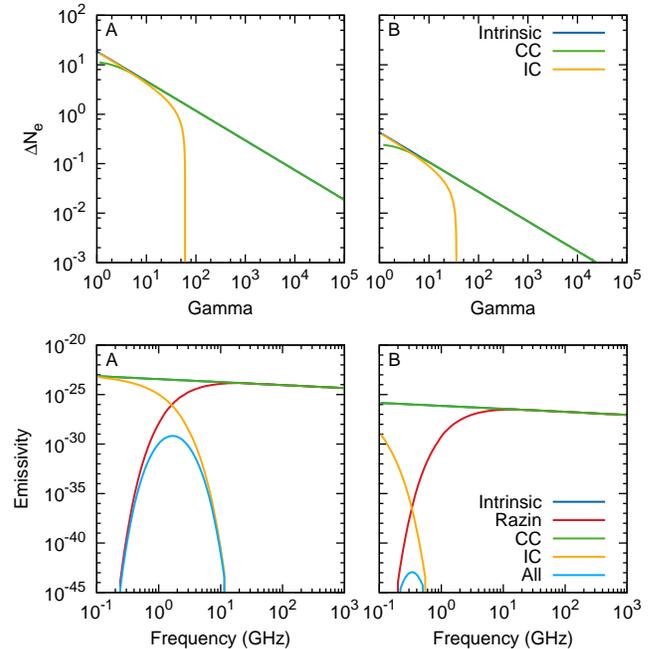}} \\
   \end{tabular}
   \caption{Relativistic electron distribution (top row) and
     emissivity (bottom row) computed from the hydrodynamic simulation
     at orbital phase, $\phi=0.8$, adopting $\zeta_{\rm rel} = 0.15$,
     $\zeta_{\rm B} = 5\times10^{-5}$, and $p=1.6$ in the model
     calculation. The left and right columns correspond to points
     ``A'' and ``B'' in Fig.~\ref{fig:ics}. Comparison of the plots
     highlights the spatial dependence of the electron distribution
     and resulting emissivity. Coulomb cooling has a relatively minor
     effect on the emissivity, hence the curve is coincident with that
     for intrinsic emission. Inverse Compton is the dominant cooling
     mechanism at the frequencies of the observations being used to
     constrain the model.}
   \label{fig:gamma_emiss}
 \end{center}
\end{figure}

To illustrate the influence of Coulomb cooling and Inverse Compton
cooling we show electron distributions and emissivities at two
representative positions (see Fig.~\ref{fig:ics}): a point close to
the apex of the wind-wind collision region (point ``A''), and a point
far downstream in the post shock flow (point ``B''). The calculation
shown in Fig.~\ref{fig:gamma_emiss} uses $\zeta_{\rm rel}=0.15$,
$\zeta_{\rm B}=5\times10^{-5}$, and $p=1.6$. In the model the total
energy in relativistic electrons scales with the gas pressure, which
is lower in the gas downstream of the shocks apex. This explains the
lower normalisation at ``B'' compared to ``A'' for the intrinsic
electron distributions in Fig.~\ref{fig:gamma_emiss}. Coulomb cooling
is responsible for a reduction at the low energy end of the electron
distribution, which equates to a turn-down in emissivity towards lower
frequencies, albeit barely noticeable in this case. However, by
comparison, the Razin effect causes a considerably larger reduction in
emissivity at low frequencies than Coulomb cooling\footnote{The
  magnitude of the reduction from the Razin effect depends on the
  magnetic field strength through the parameter $\zeta_{\rm B} \propto
  B$. If a larger value of $\zeta_{\rm B}$ were adopted, then the
  impact of the Razin effect would be smaller because the associated
  turnover frequency, $\nu_{\rm R}\propto B^{-1}$. However, the
  adopted value in Fig.~\ref{fig:gamma_emiss} is chosen based on a
  parameter space exploration, and is therefore representative of our
  ``best-fit'' value.}. There is a drastic reduction in the number of
higher energy electrons as a result of inverse Compton
cooling. Consequently, the emissivity at higher frequencies falls
considerably, and more-so in the downstream flow (``B''). Due to this
effect, the emergent radio emission essentially originates from close
to point ``A'' with only a minor contribution from regions further
downstream.

\begin{figure}
  \begin{center}
    \begin{tabular}{c}
      \resizebox{85mm}{!}{\includegraphics{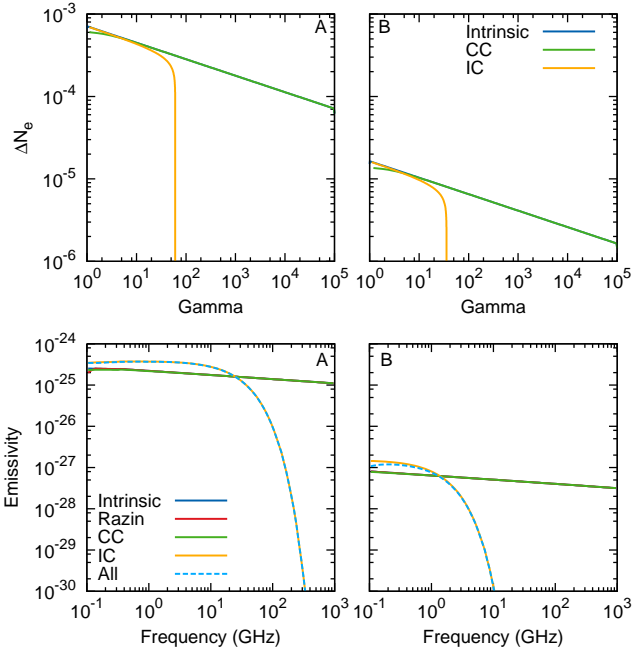}} \\
   \end{tabular}
   \caption{Relativistic electron distribution (top row) and
     emissivity (bottom row) computed from the hydrodynamic simulation
     at orbital phase, $\phi=0.8$, adopting $\zeta_{\rm rel} =
     6\times10^{-4}$, $\zeta_{\rm B} = 0.5$, and $p=1.2$. The left and
     right columns correspond to points ``A'' and ``B'' in
     Fig.~\ref{fig:ics}. Note that Inverse Compton cooling causes an
     {\it increase} in the emissivity at lower frequencies in this
     case. Also, the curves for the calculations with Coulomb cooling
     in the emissivity plots are closely with those for the intrinsic
     emission.}
   \label{fig:gamma_emiss2}
 \end{center}
\end{figure}

\begin{figure}
  \begin{center}
    \begin{tabular}{c}
      \resizebox{85mm}{!}{\includegraphics{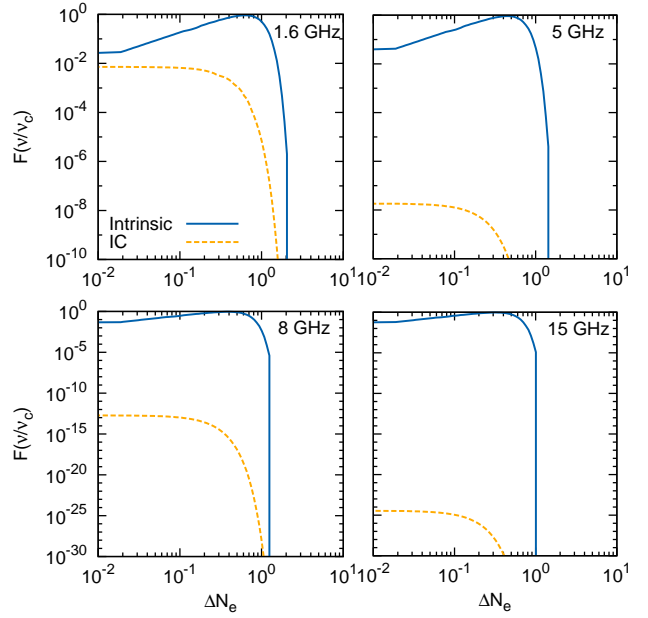}} \\
   \end{tabular}
   \caption{The dimensionless spectral shape for synchrotron emission,
     $F(\nu/\nu_c)$ (see Eq~\ref{eqn:synchrotron}) as a function of
     relativistic electron number density at orbital phase,
     $\phi=0.8$. This set of plots corresponds to a point close to the
     stagnation point (point ``A'' in Fig.~\ref{fig:ics}) with
     $\zeta_{\rm rel} = 0.15$, $\zeta_{\rm B} = 5\times10^{-5}$, and
     $p=1.6$. Results are shown for frequencies, $\nu=1.6$, 5, 8, and
     15 GHz. Corresponding electron distributions and emissivities are
     shown in the left column of Fig.~\ref{fig:gamma_emiss}; for this
     calculation, values to the left correspond to higher $\gamma$,
     and those to the right correspond to lower $\gamma$.}
   \label{fig:ne_Fnorm}
 \end{center}
\end{figure}

\begin{figure}
  \begin{center}
    \begin{tabular}{c}
      \resizebox{85mm}{!}{\includegraphics{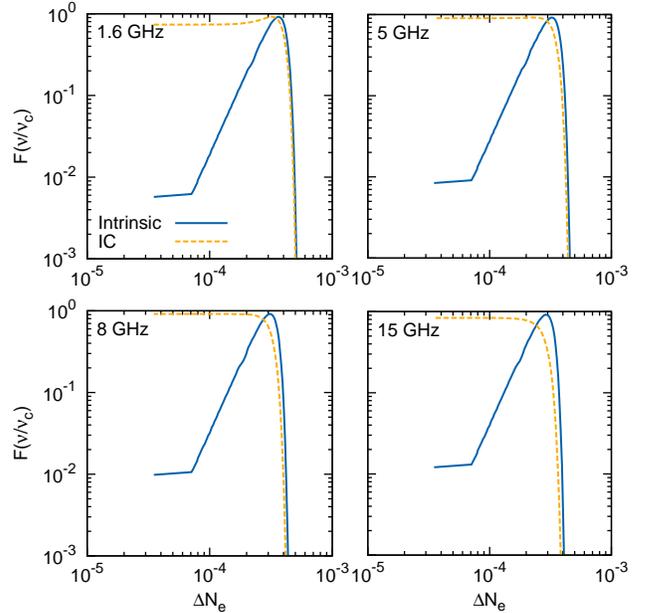}} \\
   \end{tabular}
   \caption{Same as Fig.~\ref{fig:ne_Fnorm} except for a model using
     $\zeta_{\rm rel} = 6\times10^{-4}$, $\zeta_{\rm B} = 0.5$, and
     $p=1.2$. The calculation is performed for orbital phase,
     $\phi=0.8$. The corresponding electron distribution and
     emissivities are shown in the left column of
     Fig.~\ref{fig:gamma_emiss2}.}
   \label{fig:ne_Fnorm2}
 \end{center}
\end{figure}

Repeating the above exercise for the set of parameters $\zeta_{\rm
  rel} = 6\times10^{-4}$, $\zeta_{\rm B} = 0.5$, and $p=1.2$ shows a
number of similarities and some marked differences in electron
distributions and emissivities. Examining Fig.~\ref{fig:gamma_emiss2},
one sees that Coulomb cooling has only a minor effect on the low
energy electrons, whereas inverse Compton cooling significantly
reduces the high energy electron distribution, causing an abrupt
turn-down in the relativistic electron population (in common with the
previous case). The Razin effect and Coulomb cooling have a very minor
impact on emissivities - the weaker influence of the Razin effect in
this set of calculations is the result of the considerably stronger
magnetic field which pushes the Razin-effect-related low frequency
turn-over, $\nu_{\rm R}$, to much lower frequencies. The plots of
emissivities in Fig.~\ref{fig:gamma_emiss2} also highlight an
interesting effect associated with inverse Compton cooling, namely
that the drastic cooling of the relativistic electron distribution
causes an {\it increase} in the emissivity towards lower
frequencies. The root of this increase can be traced to the change in
the characteristic frequency of synchrotron emission, $\nu_{c}$, which
affects the emissivity through the term $F(\nu/\nu_c)$ - see
Eq~\ref{eqn:synchrotron}. When inverse Compton cooling is not
included, the intrinsic electron distribution has values of $\nu/\nu_c
\ltsimm 0.3$ and in this range $F(\nu/\nu_c) \propto
(\nu/\nu_c)^{1/3}$ \citep{Rybicki:1979}. When inverse Compton cooling
is included, the relativistic electron distribution is pushed to lower
energies and there is a significant reduction in $\nu_c$. This causes
an increase in $\nu/\nu_c$, and thus $F(\nu /\nu_c)$. To illustrate
this behaviour, and the contrast in the response of $F(\nu/\nu_c)$ to
inverse Compton cooling between the cases with $\zeta_{\rm rel}=0.15$,
$\zeta_{\rm B}=5\times10^{-5}$, $p=1.6$, and with $\zeta_{\rm rel} =
6\times10^{-4}$, $\zeta_{\rm B} = 0.5$, and $p=1.2$ we show respective
plots of $F(\nu/\nu_c)$ as a function of electron number density in
Figs.~\ref{fig:ne_Fnorm} and \ref{fig:ne_Fnorm2}. Plots are shown at
frequencies, $\nu=1.6$, 5, 8, and 15 GHz. When inverse Compton cooling
is included in the case with $\zeta_{\rm rel}=0.15$, $\zeta_{\rm
  B}=5\times10^{-5}$, $p=1.6$, the result is a decrease in
$F(\nu/\nu_c)$ at all of the frequencies examined. In contrast, as
described above, for the case with $\zeta_{\rm rel} = 6\times10^{-4}$,
$\zeta_{\rm B} = 0.5$, and $p=1.2$, the inclusion of inverse Compton
cooling results in a sizeable increase in $F(\nu/\nu_c)$ across a wide
range of the electron distribution, which then causes an increase in
the emissivity.

Comparing the results for Cyg OB2$\#$9 against those for WR147 by
\cite{Pittard_et_al:2006} (see their figure 3), we note that the
smaller binary separation in the case of Cyg OB2$\#$9 leads to a more
dramatic impact of inverse Compton cooling on the emissivities -
recall from Eq~(\ref{eqn:IC}) that $\d \gamma / \d t |_{\rm IC}
\propto f_{\rm stars} \propto 1/d_{\rm sep}^2$. Indeed, in the case of
WR140 (which has $e\simeq0.9$, a comparable periastron separation to
Cyg OB2$\#$9, and roughly a factor of two larger apastron separation),
\cite{Pittard:2006} and \cite{Reimer:2006} find that inverse Compton
losses considerably diminish the population of relativistic electrons
at high energies, thus reducing the radio emission at high frequencies
for models close to periastron. These findings act to highlight the
importance of including inverse Compton cooling in models of radio
emission from massive star binaries. In particular, the result that
inverse Compton cooling can act to {\it increase} the emissivity in
cases where $\zeta_{\rm B} \gg \zeta_{\rm rel}$ has not been seen in
previous work and highlights an important, but subtle, effect.

\begin{figure}
  \begin{center}
    \begin{tabular}{c}
      \resizebox{68mm}{!}{\includegraphics{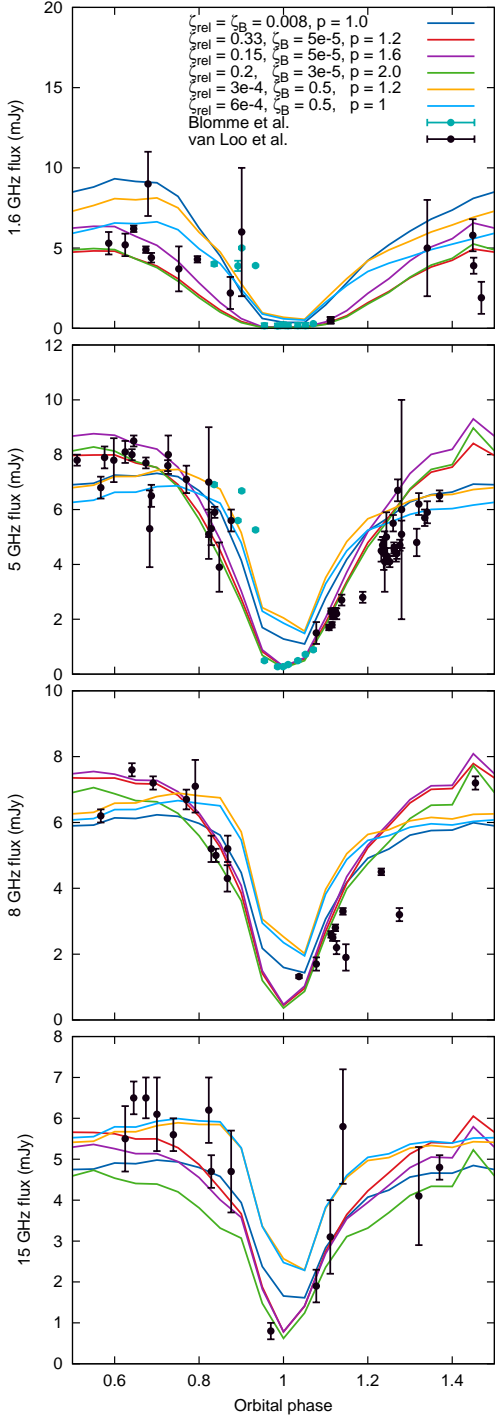}} \\
  \end{tabular}
  \caption{Radio models exploring the effect of varying $\zeta_{\rm
      rel}$, $\zeta_{\rm B}$ and $p$. The radio fluxes observed by
    \cite{vanLoo:2008} and \cite{Blomme:2013} are plotted for
    comparison.}
   \label{fig:radio_zetarel_zetaB_p_compact}
 \end{center}
\end{figure}

\begin{figure}
  \begin{center}
    \begin{tabular}{c}
      \resizebox{68mm}{!}{\includegraphics{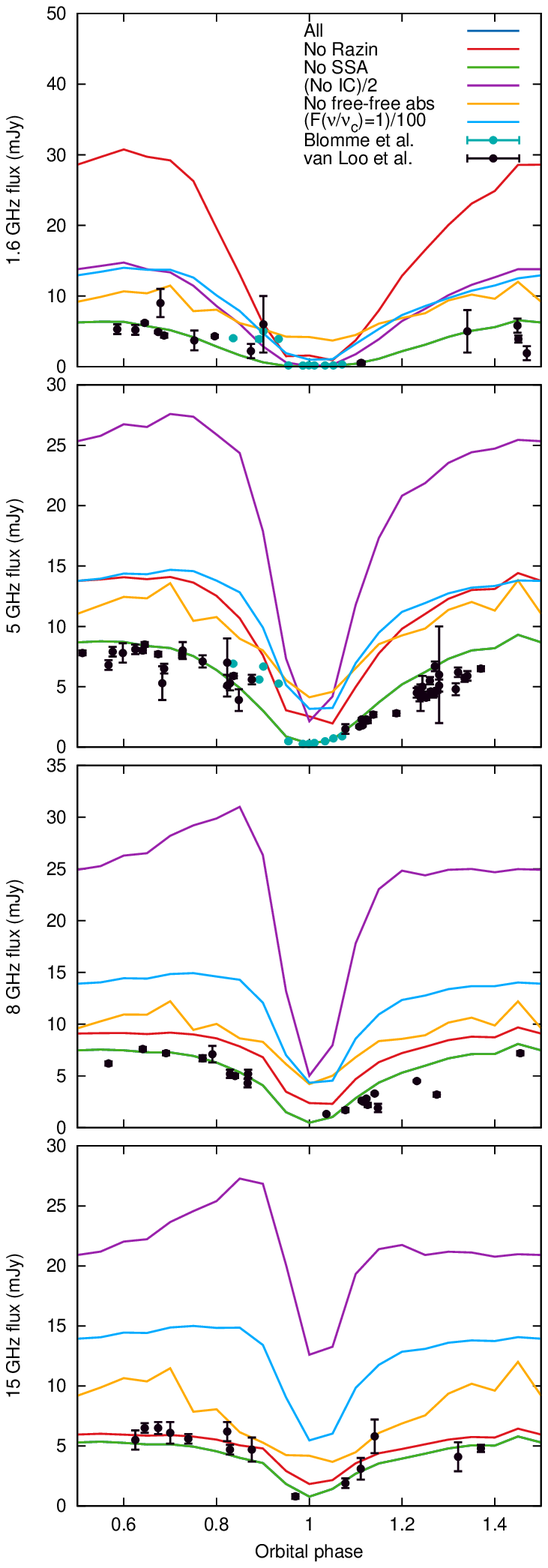}} \\
  \end{tabular}
  \caption{Radio lightcurves showing the result of removing various
    absorption/cooling effects from the model. The calculation was
    performed using $\zeta_{\rm rel}=0.15$, $\zeta_{\rm
      B}=5\times10^{-5}$, $p=1.6$, $i=60^{\circ}$, and
    $\theta=280^{\circ}$. To aid comparison, the curves for ``No IC''
    and ``$F(\nu/\nu_c) =1$'' have been rescaled downwards by factors
    of 2 and 100, respectively. The curves for ``All'' and ``No SSA''
    are closely coincident in the plots, with the latter obscuring the
    former.}
   \label{fig:radio_switches}
 \end{center}
\end{figure}

\subsection{Exploring different values of $\zeta_{\rm rel}$,
  $\zeta_{\rm B}$ and $p=2$}
\label{subsec:zetarel_zetaB_p}

The free parameters in the radio emission model are the ratios of the
energy density in relativistic electrons, $\zeta_{\rm rel}$, and in
the magnetic field, $\zeta_{\rm B}$, to the local thermal energy
density, and the power-law slope of the electron distribution
immediately post-shock, $p$. An extensive set of models were explored,
aiming to constrain these parameters. In this section we report the
key findings from the parameter space exploration.

A first approximation would be to set $\zeta_{\rm rel} =\zeta_{\rm
  B}$, such that we have a single ``injection efficiency'' parameter,
$\zeta = \zeta_{\rm rel}=\zeta_{\rm B}$, whereby magnetic and
relativistic electron energy densities are in equipartition. Such
models do, however, fail to provide a satisfactory fit to the observed
radio fluxes as the spectral slope is too steep. Indeed,
\cite{Pittard:2006} encountered similar difficulties with models
adopting $p=2$ when attempting to fit the radio spectrum of
WR\thinspace140 \citep[see also][]{Pittard_et_al:2006}.

Keeping $\zeta = \zeta_{\rm rel} = \zeta_{\rm B}$ and varying $p$ in
the range $1<p<2$ generally improves the quality of the fits, such
that $p=1$ produces a notably more satisfactory fit than
$p=2$. However, the models show an excess  radio flux at 5 GHz
compared to observations at orbital phases close to periastron
($0.95<\phi<1.05$) - see the model with $\zeta_{\rm rel} = \zeta_{\rm
  B} = 0.008$ and $p=1$ in Fig.~\ref{fig:radio_zetarel_zetaB_p_compact}.

Considering models with more degrees of freedom, such that
$\zeta_{\rm rel}$ need not be equal to $\zeta_{\rm B}$, and $p$ may be
in the range 1.2-2, a flatter radio spectrum may be produced by adopting
$\zeta_{\rm B} \ll \zeta_{\rm rel}$. In this case, by reducing
$\zeta_{\rm B}$ relative to $\zeta_{\rm rel}$ the low frequency
turn-down caused by the Razin effect rises to higher frequencies (as
$\nu_{\rm R} \propto 1/B$ - see
\citeauthor{Pittard_et_al:2006}~\citeyear{Pittard_et_al:2006} for a
further discussion of this point). The net effect is that the radio
spectrum in the 1.6-15 GHz range becomes flatter as the lower
frequency emission is diminished by the Razin effect.

A representative set of models exploring a range of values for
$\zeta_{\rm rel}$, $\zeta_{\rm B}$, and $p$ are shown in
Fig.~\ref{fig:radio_zetarel_zetaB_p_compact}. The quality of fits
improves when $p < 2$ and $\zeta_{\rm B} \ll \zeta_{\rm rel}$. For
example, a model with $\zeta_{\rm rel}=0.15$, $\zeta_{\rm B}= 5 \times
10^{-5}$, and $p=1.6$ provides a reasonable fit to the radio fluxes at
all four frequencies. Moreover, the match to the 5 GHz radio flux at
orbital phases close to periastron also improves. All models appear to
slightly over-estimate the 5 GHz flux between $1.1 < \phi < 1.3$ and
to slightly under-estimate the 15 GHz flux between $0.6 < \phi <
0.9$. The intrinsic scatter in the observational data points is also
not reproduced by the models. 

It should, however, be noted that models with $p=2$ and $\zeta_{\rm
  rel} \gg \zeta_{\rm B}$ produce as good a fit to the data as those
with, say, $p=1.2$. Hence, based on the current models there is {\it
  an indication} that $p<2$ produces valid fluxes, yet a firm
distinction cannot be made. Another factor that is apparent from
Fig.~\ref{fig:radio_zetarel_zetaB_p_compact} is that some of the
models require relatively large values of $\zeta_{\rm rel}$ of a few
tens of percent, suggesting efficient particle acceleration
in Cyg OB2$\#$9. Such large values of $\zeta_{\rm rel}$ raise the
question of whether shock modification may be causing values of
$p<2$. We return to this point at the end of \S~\ref{sec:conclusions}.

A further examination of parameter space reveals that reasonable fits
to the radio data can be acquired for a set of models that have $
\zeta_{\rm rel} \ll \zeta_{\rm B}$ and $1<p<1.2$
(Fig.~\ref{fig:radio_zetarel_zetaB_p_compact}). Models of this type
require a power-law slope closer to unity (i.e. $p \ll 2$) to obtain a
suitably flat spectrum. In addition, values of $\zeta_{\rm B}$ of
order a few tens of percent are preferred by the fits to
observations. A notable weakness of this set of models is the excess
in the 5 and 8 GHz fluxes at phases close to periastron. Furthermore,
compared to models with $ \zeta_{\rm rel} \gg \zeta_{\rm B}$, the
models with $\zeta_{\rm rel} \ll \zeta_{\rm B}$ provide a somewhat
poorer quality fit to the rise of the 5 and 8 GHz fluxes between
phases, $1.1 \leq \phi \leq 1.4$. For instance, a model with
$\zeta_{\rm rel}= 3\times10^{-4}$, $\zeta_{\rm B}= 0.5$, and $p=1.2$
(orange curve) provides arguably the best fit to observations of the
curves, yet it displays excess flux at phases close to periastron at
5, 8, and 15 GHz.

In summary, model fits to the observed radio fluxes of arguably
similar quality can be attained with either $\zeta_{\rm rel} \gg
\zeta_{\rm B}$ or $\zeta_{\rm rel} \ll \zeta_{\rm B}$, with some
suggestion of better quality fits when $p <2$ is adopted. Therefore,
based on the current models, firm constraints cannot be placed on
$\zeta_{\rm rel}$, $\zeta_{\rm B}$, and $p$, although the models do
provide some guidance on the regions of parameter space of
interest. For example, $\zeta_{\rm rel}=0.15$, $\zeta_{\rm B}= 5\times
10^{-5}$, and $p=1.6$, or, $\zeta_{\rm rel}= 3\times10^{-4}$,
$\zeta_{\rm B}= 0.5$, and $p=1.2$.

\subsection{Dissecting contributions to the radio emission}
\label{subsec:radio_contributions}

To give some indication of the physics causing the variation in the
radio emission as a function of orbital phase, in this section we
examine the separate contributions from the Razin effect, synchrotron
self absorption, free-free absorption, inverse Compton cooling, and
the function $F(\nu/\nu_c)$ (which determines the spectral shape - see
Eq~\ref{eqn:synchrotron}). To establish whether different combinations
of these effects can explain the differences between models with
$\zeta_{\rm rel} \ll \zeta_{\rm B}$ and $\zeta_{\rm rel} \gg
\zeta_{\rm B}$ this exercise is repeated for an example case from both
sets of models. In Fig.~\ref{fig:radio_switches} the influence of
individually removing the various attenuation/cooling mechanisms on
the radio lightcurves is shown for a model with $\zeta_{\rm
  rel}=0.15$, $\zeta_{\rm B}=5\times10^{-5}$, and $p=1.6$ - the
``standard'' case including all attenuation/cooling mechanisms is
represented by the dark blue curve (which is essentially coincident
with the green curve). Note that the larger the difference between the
standard case and a respective curve, the greater the magnitude of
that effect. The Razin effect is largest at low frequencies and
decreases towards higher frequencies, illustrated by the difference
between the curves for ``No Razin'' and ``All'' being smaller for the
higher frequency fluxes. Similarly, inverse Compton cooling affects
the higher frequency flux more-so than that at lower frequencies,
consistent with the results in \S~\ref{subsec:IC}. Synchrotron self
absorption is essentially negligible.

It is evident from the various curves shown in
Fig.~\ref{fig:radio_switches} that a minimum exists in all cases,
albeit less pronounced in the case with, for example, no free-free
absorption. This fact can be further demonstrated by setting
$F(\nu/\nu_c)=1$, which removes the dependence of synchrotron emission
on the Lorentz factor ($\nu_c \propto \gamma^2$ - see
Eqns~\ref{eqn:synchrotron} and \ref{eqn:nu_char}, and
\S~\ref{subsec:IC}). As such, when $F(\nu/\nu_c)=1$, the total
synchrotron flux, $P_{\rm syn-tot}(= n_{\rm e-rel} P_{\rm syn})$,
depends on the magnetic field strength and the number density of
relativistic electrons, $P_{\rm syn-tot} \propto n_{\rm e-rel}
B$. From Fig.~\ref{fig:radio_switches} one can see a clear minimum in
this case. $n_{\rm e-rel}$ depends on the pre-shock ram pressure via
$n_{\rm e-rel} \propto \zeta_{\rm rel} \rho v_{\rm pre}^2 \propto
\zeta_{\rm rel} d_{\rm sep}^{-2} v_{\rm pre}$. Similarly, the magnetic
field scales as $B\propto (\zeta_{\rm B} v_{\rm pre})^{1/2} d_{\rm
  sep}^{-1}$. Combining these scalings gives a relation for the
volume-integrated\footnote{The volume of the emitting region scales as
  $d_{\rm sep}^{3}$.} synchrotron power (when $F(\nu/\nu_c)=1$) of
$P_{\rm syn-tot} \propto \zeta_{\rm rel} \zeta_{\rm B}^{1/2} v_{\rm
  pre}^{3/2}$. Note that this latter relation is independent of the
binary separation and only scales with the pre-shock wind
velocity. This explains the flatness of the ``$F(\nu/\nu_c)$=1'' curve
at phases away from periastron when $v_{\rm pre} $ is roughly constant
(Fig.~\ref{fig:radio_switches}), and the dip into a minimum which
coincides with the variation in $v_{\rm pre}$
(Fig.~\ref{fig:vpre_chi}).

The above relations can also be extended to understand the behaviour
of the calculation without inverse Compton cooling (``No IC'' in
Fig.~\ref{fig:radio_switches}). In this case the Lorentz factors for
the relativistic electrons are largely un-affected
(\S~\ref{subsec:IC}). Noting that $\nu/ \nu_c \gg 1$, such that to
first-order\footnote{When $\nu/\nu_c \gg 1$, $F(\nu/\nu_c) \propto
  (\nu/\nu_c)^{1/2} \exp(\nu/\nu_c)$ \citep{Rybicki:1979}. Expanding
  the exponential term to first order, $\exp(\nu\nu_c) \approx 1 +
  \nu/\nu_c \sim \nu/\nu_c$ when $\nu/\nu_c \gg 1$, leads to
  $F(\nu/\nu_c) \propto (\nu/\nu_c)^{3/2}$.}  $F(\nu/\nu_{c}) \propto
(\nu/\nu_c)^{3/2}$ \citep{Rybicki:1979}, the relation for the
synchrotron power becomes $P_{\rm syn-tot} \propto \zeta_{\rm rel}
\zeta_{\rm B}^{-1/4} v_{\rm pre}^{9/4} d_{\rm sep}^{3/2}$. The
introduction of a dependence on binary separation and the stronger
scaling with pre-shock wind velocity explain the larger phase
dependence of the radio flux for the ``No IC'' curve (compared to that
for $F(\nu/\nu_c)$ considered in the previous paragraph). The model
fluxes vary by factors of 100, 31, 15, and 6.6 at 1.6, 5, 8, and 15
GHz, respectively, which should be compared against the scaling
relation for $P_{\rm syn-tot}$ given above, which yields a ratio of
$\simeq 21$. Hence, the variation in flux at 8 and 15 GHz can be
explained as the result of changes in the post-shock gas conditions,
mostly linked to the binary separation. The 1.6 and 5 GHz fluxes, on
the other hand, require some additional flux variation. The leading
candidates are free-free absorption, inverse Compton cooling, and the
Razin effect (see Fig.~\ref{fig:radio_switches}).

We may therefore conclude that the cause of the minimum for the model
with $\zeta_{\rm rel}=0.15$, $\zeta_{\rm B}=5\times10^{-5}$, and
$p=1.6$ is a combination of factors. Firstly, free-free absorption
plays a role because the minimum in the lightcurves close to
periastron ($\phi=1$) is less pronounced when it is not
included. Second, the Razin effect contributes to the depth of the
minimum because, when it is not included, the minimum has a flatter
bottom. Thirdly, as described above, the synchrotron emission depends
on the pre-shock wind velocities through the post-shock gas
pressure. Thus, a reduction in the pre-shock wind speed close to
periastron will cause a relative decline in the number density of
relativistic electrons, and also the magnetic field strength, given
the scaling in our model. Indeed, the flux variations in
Fig.~\ref{fig:radio_switches} are coincident with the change in
pre-shock wind speeds in Fig.~\ref{fig:vpre_chi}.

\begin{figure}
  \begin{center}
    \begin{tabular}{c}
      \resizebox{68mm}{!}{\includegraphics{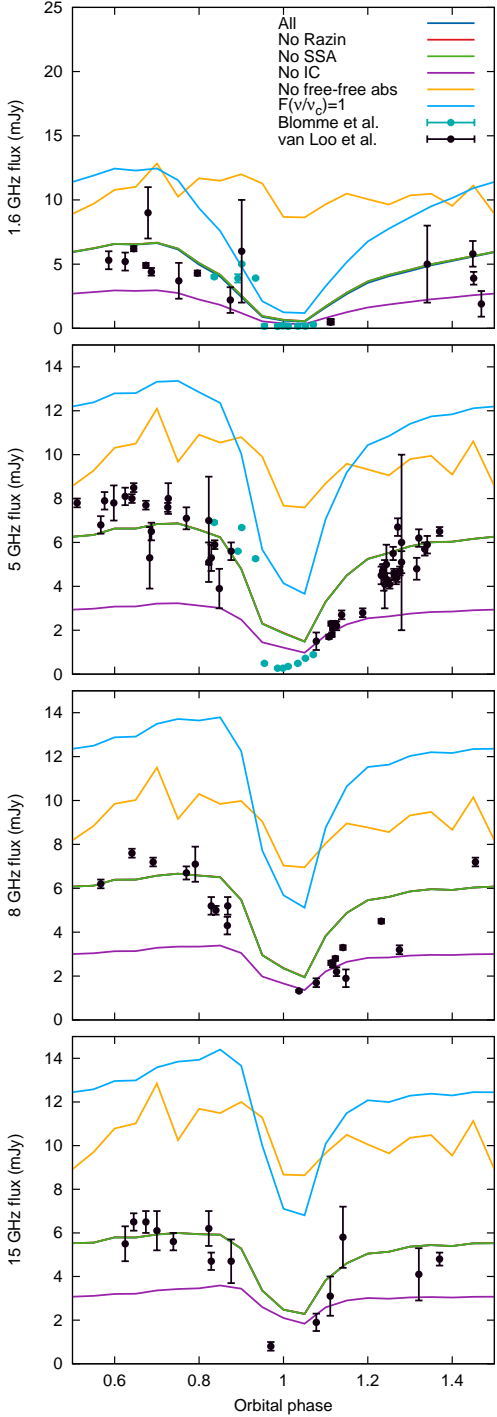}} \\
  \end{tabular}
  \caption{Same as Fig.~\ref{fig:radio_switches} except for
    $\zeta_{\rm rel}=3\times10^{-4}$, $\zeta_{\rm B}=0.5$, and
    $p=1.2$. The curves for ``All'', ``No SSA'', and ``No Razin''' are
    closely coincident in the plots.}
   \label{fig:radio_switches2}
 \end{center}
\end{figure}

Next we turn to Fig.~\ref{fig:radio_switches2} which shows the case
with $\zeta_{\rm rel}=3\times10^{-4}$, $\zeta_{\rm B}=0.5$, and
$p=1.2$. The stronger magnetic field strength in this case renders the
Razin effect less important (see \S~\ref{subsec:IC}). The largest
reduction in emission in this case is from free-free
absorption. Interestingly, including inverse Compton cooling actually
leads to an {\it increase} in emission. The reasons for this relate to
the decrease in the characteristic frequency for synchrotron emission
that is caused by the cooling of the relativistic electron
distribution, and the shape of the function $F(\nu/\nu_c)$ that
described the emitted synchrotron spectrum - see \S~\ref{subsec:IC}
for further discussion. Examining the scaling of the synchrotron power
with no inverse Compton cooling, and when $\nu/\nu_c \ll 1$ such that
$F(\nu/\nu_c)\propto (\nu/\nu_c)^{1/3}$, one finds $P_{\rm syn-tot}
\propto \zeta_{\rm rel} \zeta_{\rm B}^{1/3} v_{\rm pre}^{10/6} d_{\rm
  sep}^{1/3}$. The weaker scaling with binary separation explains the
shallower minimum for the ``No IC'' curve in
Fig.~\ref{fig:radio_switches2} compared to
Fig.~\ref{fig:radio_switches}. Removing the influence of $\nu_c$ (by
setting $F(\nu/\nu_c)=1$ - light blue curve in
Fig.~\ref{fig:radio_switches2}), the normalisation of the lightcurve
rises (compared to the standard case). The fact that a minimum is
still observed when $F(\nu/\nu_c)=1$ again indicates that flux
variations must be related to the post-shock gas pressure, in a
similar way as to the case with $\zeta_{\rm rel}=0.15$, $\zeta_{\rm
  B}=5\times10^{-5}$, and $p=1.6$ described in the preceding
paragraphs. The ratio of apastron and periastron radio fluxes for the
model with $\zeta_{\rm rel}=3\times10^{-4}$, $\zeta_{\rm B}=0.5$, and
$p=1.2$ at 1.6, 5, 8, and 15 GHz are 10.7, 3.3, 2.5, and
2.1. Evaluating the relation for $P_{\rm syn-tot}$ earlier in this
paragraph, the estimated variation in the synchrotron flux due to
changes in the post-shock gas conditions between apastron and
periastron is $\simeq 2.4$. Therefore, the flux variations at 8 and 15
GHz are consistent with intrinsic changes to the shocks, whereas the
lower frequency fluxes leave room for additional absorption/cooling
mechanisms, namely free-free absorption and inverse Compton
cooling. Indeed, without free-free absorption the lightcurves are
notably flatter with a less pronounced minimum, and, when inverse
Compton cooling is included the flux at phases away from periastron
($\phi \ltsimm 0.9$ and $\phi \gtsimm 1.1$) is higher, making the
minimum more pronounced.

\begin{figure}
  \begin{center}
    \begin{tabular}{c}
      \resizebox{80mm}{!}{\includegraphics{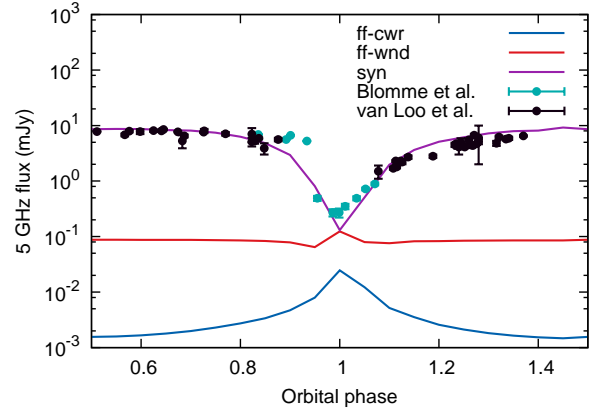}} \\
  \end{tabular}
  \caption{Radio lightcurve at 5GHz showing the separate contributions
    to the emission: free-free emission from the colliding winds
    region (ff-cwr), free-free emission from the un-shocked winds
    (ff-wnd), and synchrotron emission from the colliding winds region
    (syn). The calculation was performed using $\zeta_{\rm rel}=0.15$,
    $\zeta_{\rm B}=5\times10^{-5}$, $p=1.6$, $i=60^{\circ}$, and
    $\theta=280^{\circ}$. Non-thermal emission clearly dominates at
    radio wavelengths for Cyg OB2$\#$9.}
   \label{fig:radio_ffsyn}
 \end{center}
\end{figure}

Finally, before closing this section we examine the separate
contributions to the radio flux from free-free and synchrotron
emission. As shown in Fig.~\ref{fig:radio_ffsyn}, synchrotron emission
dominates throughout the majority of the orbit. The contribution from
free-free emission is noticeably lower, with the wind-wind collision
region contributing a negligible flux, consistent with the model
estimates by \cite{Blomme:2013}. However, at periastron the
synchrotron flux falls sharply and reaches a comparable level to the
free-free emission from the un-shocked winds. Scaling the un-shocked
winds free-free flux estimates from \cite{Blomme:2013} to account for
the revised mass-loss rates derived in the present work, gives a value
of 0.034 mJy (for both winds), noticeably lower than the value of
$\simeq 0.065-0.12\;$mJy in the model calculation
(Fig.~\ref{fig:radio_ffsyn}). The explanation for the additional flux
comes from mixing between the shocked and un-shocked winds which
increases the free-free flux from the former. Moreover, this also
explains the rise in the free-free flux from the un-shocked winds at
phases close to periastron, occurring as a result of the increased
level of turbulent mixing at these phases. From
Fig.~\ref{fig:radio_ffsyn} we conclude that Cyg OB2$\#$9 is
predominantly a non-thermal radio emitter with free-free emission from
the un-shocked stellar winds becoming important only for a very short
interval close to periastron.

\subsection{Spectral index}
\label{subsec:radio_index}

\begin{figure}
  \begin{center}
    \begin{tabular}{c}
      \resizebox{80mm}{!}{\includegraphics{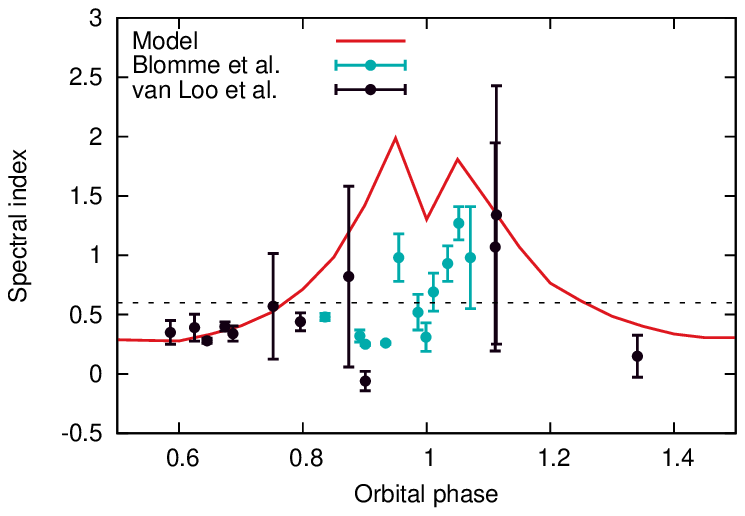}} \\
  \end{tabular}
  \caption{Spectral index between 1.6 and 5 GHz as a function of
    orbital phase. The calculation was performed using $\zeta_{\rm
      rel}=0.15$, $\zeta_{\rm B}=5\times10^{-5}$, $p=1.6$,
    $i=60^{\circ}$, and $\theta=280^{\circ}$. Observationally inferred
    spectral index values from \cite{vanLoo:2008} and
    \cite{Blomme:2013} are shown for comparison - error bars have been
    computed by error propagation. The dashed horizontal line at
    spectral index of 0.6 corresponds to the canonical value for
    free-free emission from a single massive star wind.}
   \label{fig:radio_index}
 \end{center}
\end{figure}

In Fig.~\ref{fig:radio_index} we show the spectral index\footnote{The
  spectral index, $s$, is the exponent in the relation $F_{\nu}
  \propto \nu^s$. For both model and observational data, the spectral
  index was computed using radio data at 1.6 and 5 GHz.} computed from
a model with $\zeta_{\rm rel}=0.15$, $\zeta_{\rm B}=5\times10^{-5}$,
$p=1.6$, $i=60^{\circ}$, and $\theta=280^{\circ}$, compared to
observationally inferred values from \cite{vanLoo:2008} and
\cite{Blomme:2013}. At orbital phases away from periastron ($\phi
\ltsimm 0.75$ and $\phi \gtsimm 1.25$) the model spectral index is in
good agreement with the data points from \cite{vanLoo:2008}. The fact
that the spectral index is lower than the canonical value for
free-free emission from single star winds ($s=0.6$ -
\citeauthor{Wright:1975}~\citeyear{Wright:1975}) highlights the
apparent non-thermal emission. However, the model and observations
diverge somewhat at orbital phases closer to periastron ($0.75 < \phi
< 1.25$). This disagreement follows from the poorer fit of the model
to observations at 1.6 and 5 GHz at these phases (see
Fig.~\ref{fig:radio_zetarel_zetaB_p_compact}). Encouragingly, the
model does reproduce the general trend observed by \cite{Blomme:2013}
of a rise in spectral index at $\phi \sim 0.95$, followed by a sharp
fall around $\phi \sim 1$, and then a second rise at $\phi \sim
1.05$. It should be noted that at phases close to periastron ($0.75 <
\phi < 1.25$) synchrotron emission dominates over free-free
(Fig.~\ref{fig:radio_ffsyn}), however the spectral index inferred from
the 1.6 and 5 GHz fluxes indicates a slope $\gg 0.6$ which would
typically be taken as evidence for {\it thermal} emission. Examining
the curves with and without the Razin effect in
Fig.~\ref{fig:radio_switches}, the rise in spectral index approaching
periastron is the result of a larger decrease, and earlier onset of
decline, in flux at 1.6 GHz than at 5 GHz due to the Razin effect.

\section{Discussion}
\label{sec:discussion}

\subsection{Deeper insight from multi-epoch analysis}

When performing model fits for radio emission from WR\thinspace140,
\cite{Pittard:2006} found that favourable fits to the observed
spectrum could be acquired for two distinct families of values for the
parameters $\zeta_{\rm rel}$, $\zeta_{\rm B}$, and $p$. As such, from
their fits to the radio spectrum at a single orbital phase, they
concluded that there was a degeneracy in their parameter space
exploration. One set of best-fit parameters was localised around large
values of $\zeta_{\rm rel}$, a large ratio between $\zeta_{\rm rel}$
and $\zeta_{\rm B}$, and a value of $p$ in the range 1.6-2. The second
set of models favoured values of $p \ltsimm 1.4$, smaller $\zeta_{\rm
  rel}$, and $\zeta_{\rm B} \gg \zeta_{\rm rel}$. Indeed,
\cite{Pittard:2006} note a degeneracy between $\zeta_{\rm B}$ and
$p$. The models considered for Cyg OB2$\#$9 in
\S\S~\ref{subsec:zetarel_zetaB_p} (see
Fig.~\ref{fig:radio_zetarel_zetaB_p_compact}) are representative of
the two families detailed by \cite{Pittard:2006} for
WR\thinspace140. Interestingly, by fitting models to the radio data
covering the entire orbit, as has been done in this work,
distinguishing features between the two families of models become
apparent. For instance, based on a comparison of models in
Fig.~\ref{fig:radio_zetarel_zetaB_p_compact}, models with $\zeta_{\rm
  rel} \gg \zeta_{\rm B}$ provide a better fit to the 5, 8, and 15 GHz
radio data at phases close to periastron ($\phi \simeq 1$). In
contrast, models with $\zeta_{\rm rel} \ll \zeta_{\rm B}$ have an
excess flux at phases close to periastron and too quick a post-minimum
recovery. However, as noted in \S~\ref{subsec:zetarel_zetaB_p}, the
overall quality of the fits from the two families of parameters
(i.e. $\zeta_{\rm rel} \gg \zeta_{\rm B}$ or $\zeta_{\rm rel} \ll
\zeta_{\rm B}$) is arguably similar, and neither emerges as a firm
favourite. Yet the models do demonstrate that fits to multi-epoch data
have the potential to provide deeper insight from contrasting
behaviour at different orbital phases, with the hope that future, more
sophisticated models will be able to place stronger constraints on
model parameters.

\subsection{Constraining the viewing angle}
\label{subsec:los}

There is a correspondence between the adopted stellar masses
and the inclination angle, $i$, of the binary orbit which results from
the radial velocity analysis used to derive the orbital
solution. Hence, it is important to clarify whether the results are
strongly dependent on the adopted inclination angle. For our adopted
stellar masses we have $i=60^{\circ}$, consistent with previous
estimates by \cite{vanLoo:2008}. The second angle of interest when
defining the binary orientation is the angle subtended between the
line-of-centres and the line-of-sight, $\theta$, (projected onto the
orbital plane). The analysis by \cite{Naze:2012} determined $\theta
\simeq 280^{\circ}$. 

Supplementary tests, adopting ``best-fit'' parameters from the X-ray
analysis (\S~\ref{sec:xrays}) and $\zeta_{\rm rel}=0.15$, $\zeta_{\rm
  B}= 5\times 10^{-5}$, and $p=1.6$ (\S~\ref{subsec:zetarel_zetaB_p}),
and where $i$ and $\theta$ were varied, show little dependence of the
X-ray or radio lightcurves on viewing angle, as might be expected given
the mass-loss rates adopted. As such, the observed radio and X-ray
fluxes cannot be used to constrain the orientation of the system. This
may, of course, change in future models where the synchrotron
emission/absorption is {\it not} assumed to be isotropic
\citep[e.g.][]{Reimer:2006, Reitberger:2014}.

\subsection{Clumpy winds}
\label{subsec:clumpy_winds}

In the present study we have assumed that the winds are
smooth. However, there is considerable theoretical and observational
evidence for clumpy stellar winds \citep[see, e.g.,][]{Puls:2008}. As
the shocks are quasi-adiabatic throughout the orbit clumps, if present
in the un-shocked winds, will be rapidly destroyed in the post-shock
region \citep{Pittard:2007}. Hence, we may safely assume that the
thermal X-rays and non-thermal (synchrotron) radio emission would not
be affected by clumpy stellar winds. The thermal (free-free) radio
emission and absorption does, however, change for clumpy winds. A
clumpy wind mimics the emission of a higher mass-loss rate wind,
i.e. $\dot{M}_{\rm clumpy} = \dot{M}_{\rm actual} f_{\rm
  fill}^{-1/2}$, where $f_{\rm fill}$ is the filling-factor for an
inhomogeneous wind \citep{Lamers:1984} - see also the discussion in
\cite{Pittard_et_al:2006}. As such, the net effect of clumpy winds is
an increase in the free-free emission by a factor $f_{\rm
  fill}^{-2/3}$. The results presented in
\S~\ref{subsec:radio_contributions} showed that the free-free emission
from the (smooth) un-shocked winds is roughly two orders of magnitude
fainter than the synchrotron flux for a large part of the orbit, with
a comparable flux level achieved for only a brief interval close to
periastron (see Fig.~\ref{fig:radio_ffsyn}). An increase in flux by a
factor of one hundred equates to a clumping factor ($=1/f_{\rm fill}$)
of one thousand, well beyond current estimates for hot star winds. In
contrast, at phases close to periastron a relatively modest clumping
factor (e.g. $f_{\rm fill} \gtsimm 3$) could cause the free-free
emission from the un-shocked winds to imprint significantly on the
total emergent flux.

Clumping also may also have an important influence on absorption. For
instance, for a smooth wind we determine the radius of optical depth
unity \citep[using the expression from][]{Wright:1975}, $R_{\tau=1} =
1359$, 637, 466, and 306$\;R_{\odot}$ at frequencies of 1.6, 5, 8, and
15 GHz, respectively. These estimates increase by a factor of $f_{\rm
  fill}^{-1/3}$ when wind inhomogeneity is accounted for due to the
increase in free-free absorption increasing. Therefore, if the winds
are clumpy, the synchrotron flux may be subject to larger
attenuation. This effect has the potential to impact on the shape of
the radio lightcurves, particularly close to periastron where lines of
sight to the apex of the CWR pass deeper into the un-shocked winds. It
would, therefore, be of interest to examine the effect of wind
clumping in future models of Cyg OB2$\#$9.

\subsection{Surface magnetic field estimate}

The radio emission models presented in \S~\ref{sec:radio} provide
valuable insight into the quantitative details of particle
acceleration occurring at the wind-wind collision shocks. One of the
parameters that is returned by the models is the fraction of thermal
energy in the magnetic field, $\zeta_{\rm B}$. Using some simple
assumptions regarding the radial dependence of the magnetic fields, we
may use $\zeta_{\rm B}$ to estimate the surface magnetic field
strengths of the stars. For completeness we will derive estimates for
non-rotating stars (with radial magnetic fields) and rotating stars
(where the magnetic field is toroidal). Let us consider the post-shock
region along the line-of-centres, which is roughly equidistant from
both stars, and begin with the non-rotating stars case. Assuming a
priori that the magnetic field is of equipartition strength (or
weaker), field lines will be drawn out by the wind to have a radial
configuration \citep[see, for
example,][]{Falceta-Goncalves:2012}. Therefore, the magnetic field
close to the line-of-centres will be (almost) parallel to the shock
normal and, due to the divergence-free constraint, there will be no
change in the magnetic field across the shock. If we then assume that
the Alfv{\' e}n radius is coincident with the stellar surface, the
radial dependence of the magnetic field will vary as $\propto
(R_{\ast}/r)^2$ \citep{Eichler_Usov:1993}. Combining the above, we may
estimate the surface magnetic field strength on the stars from,
$B_{\rm surf} \approx (r/R_{\ast})^2 [(4 \pi/3) P \zeta_{\rm
  B}]^{1/2}$. Inserting a typical value from the radio analysis of
$\zeta_{\rm B}= 5\times10^{-5}$ and using the hydrodynamic simulation
to evaluate the remaining terms, we obtain estimates of $B_{\rm
  surf}\simeq 8-52\;$G, where the range of values arises from
estimates at periastron and apastron, respectively.

Repeating the above exercise for the case of (slowly) rotating stars
($v_{\rm rot}/v_{\infty} = 0.1$ where $v_{\rm rot}$ is the surface
rotation velocity of the stars), the radial dependence of the toroidal
magnetic field goes as $B = B_{\rm surf} (v_{\rm rot}/v_{\infty})
(R_{\ast}/r)$ \citep{Eichler_Usov:1993}. Also, the magnetic field will
be close-to-perpendicular to the shock, resulting in an enhancement in
the post-shock magnetic field strength by a factor of $\sim 4$
\citep{Draine:1993}. The expression for the surface magnetic field
then becomes, $B_{\rm surf} \approx 5.12 (r/R_{\ast})(P \zeta_{\rm
  B})^{1/2}$. Inserting appropriate values gives $B_{\rm surf} \simeq
0.3-1.7\;$G. Hence, in the limit that the stars are rotating, albeit
slowly, we arrive at estimates of slightly weaker surface magnetic
field strengths because the magnetic field falls-off with radius at a
slower rate.

The estimated values of $B_{\rm surf}$ using $\zeta_{\rm
  B}=5\times10^{-5}$ are consistent with weakly magnetized massive
stars, lending support to our radio fits. Furthermore, they are lower
than magnetic field strengths inferred for strongly magnetized, hard
X-ray emitting single stars \citep[][]{Naze:2012b}. In contrast, if we
were instead to adopt $\zeta_{\rm B}=0.5$ in our $B_{\rm surf}$
calculations (to be consistent with the $\zeta_{\rm B} \gg \zeta_{\rm
  rel}$ models), the estimates shift to $B_{\rm surf} \simeq
30-5200\;$G. Towards the upper limit of this more highly magnetized
regime, the influence of magnetic fields on wind dynamics would become
important \citep[e.g.][]{ud-Doula:2002}. However, the above
calculations do not consider the possibility of magnetic field
amplification \citep[e.g.][]{Lucek:2000, Bell:2001}, which could
reduce the estimated surface magnetic field strengths.

\subsection{Future directions}

In the present investigation of Cyg OB2$\#$9 we have extensively
studied the hydrodynamics of the wind-wind collision and the emergent
X-ray and radio emission. We have used a high-resolution,
three-dimensional hydrodynamic simulation to gain a good description
of the density and temperature distributions in the system as a
function of orbital phase. Explorations of additional effects such as
differing electron and ion temperatures, mass-loss rates, and
efficiencies of particle acceleration have been treated separately
from the hydrodynamics, such that variations in the magnitude of these
effects are not fed back into the gas dynamics. The major advantage of
the approach adopted in this work is that a thorough parameter space
exploration has been possible because we have tailored the
calculations to reduce the computational strain. The benefit of this
approach is that future studies of Cyg OB2 $\#$9 can focus on a much
narrower range of parameter space, and can concentrate on advancing
the physics prescriptions. In particular, examining different
wind-momentum ratios may produce a more asymmetric minimum in the
radio lightcurves, as is required to improve the agreement of the
models to observed data points - see the discussion by
\cite{Blomme:2013}.

With future studies in mind, it would be useful to perform simulations
in this narrower parameter space range that self-consistently include
the effect of variations in, for example, the initial post-shock
electron and ion temperatures on the hydrodynamics. For similar
reasons, it would be desirable to have simulations which treat Coulomb
and inverse Compton cooling self-consistently with the dynamics
\citep[e.g.,][]{Pittard_et_al:2006, Pittard:2006}, include magnetic
fields \citep[e.g.][]{Falceta-Goncalves:2012}, account for anisotropic
synchrotron emission, and treat inverse Compton scattering in the
Klein-Nishina limit \citep[e.g.][]{Reimer:2006}. Ideally, a preferred
approach would solve the cosmic ray transport equation directly with
the hydrodynamics \citep[e.g.][]{Reitberger:2014}. Moreover, treating
the variation in post-shock heating for multiple species would permit
a more complete analysis of non-equilibrium ionisation
\citep[e.g.][]{Pollock:2005, Zhekov:2007}. A more self-consistent
approach would also make possible models of systems where the CWR is
turbulent.

\section{Conclusions}
\label{sec:conclusions}

We have performed the first detailed study of wind-wind collision
dynamics and X-ray and radio emission utilising three-dimensional
hydrodynamic models and radiative transfer, with an analysis of
multiple epoch data. The wind-wind collision dynamics of Cyg OB2$\#$9
has been modelled using a three dimensional adaptive-mesh refinement
(AMR) simulation (including wind acceleration, radiative cooling, and
orbital motion), which then acts as the input to radiative transfer
calculations for the emergent X-ray and radio emission. Our
main results may be summarised as follows:
\begin{itemize}
\item The alteration to wind acceleration along the line-of-centres
  caused by the interacting stellar radiation fields differs from the
  estimates from 1D static star calculations. For the primary star we
  observe a sharp fall in the pre-shock wind speed close to
  periastron, whereas there is a small rise in the companion's
  pre-shock wind speed. 
\item The X-ray emission analysis presents strong evidence for a low
  ratio of electron-to-ion temperatures in the immediately post-shock
  gas, and a substantial lowering of mass-loss rates compared to
  previous estimates. The observations can be matched well using a
  model with an initial post-shock electron-to-ion temperature ratio
  of $\tau_{\rm e-i} = T_{\rm e}/T|_0 = 0.1$, and mass-loss rates of
  $\dot{M}_1= 6.5\times 10^{-7}\;{\rm M_{\odot}~yr^{-1}}$ and
  $\dot{M}_2= 7.5\times 10^{-7}\;{\rm M_{\odot}~yr^{-1}}$. An
  examination of X-ray spectra reveals the models to marginally
  over-estimate the observed flux at both the lowest and highest
  energies.
\item Both the X-ray and radio emission are essentially viewing angle
  independent.
\item The results of a detailed radio model parameter space
  exploration for $\zeta_{\rm rel}$, $\zeta_{\rm B}$, and $p$ show
  that models which assume equipartition between the magnetic field
  and relativistic electron energy densities ($\zeta = \zeta_{\rm
    B}=\zeta_{\rm rel}$) fail to reproduce the observations, even when
  the slope of the relativistic electron distribution is allowed to
  vary in the range $1<p<2$ (see \S~\ref{subsec:zetarel_zetaB_p}). The
  agreement of the models with observations improves when
  equipartition is relaxed. However, models of arguably similar
  fit-quality are produced by contrasting sets of parameters, namely
  $\zeta_{\rm rel} \gg \zeta_{\rm B}$ and $\zeta_{\rm rel} \ll
  \zeta_{\rm B}$. Example values that provide reasonable fits are
  $\zeta_{\rm rel}=0.15$, $\zeta_{\rm B}=5\times10^{-5}$, and $p=
  1.6$, and, $\zeta_{\rm rel}=3 \times 10^{-4}$, $\zeta_{\rm B}=0.5$,
  and $p = 1.2$.
\item The radio calculations for Cyg OB2$\#$9 show that non-thermal emission
clearly dominates over the thermal contribution (see
Fig.~\ref{fig:radio_ffsyn}) confirming the prediction from
\cite{Blomme:2013}. 
\item The variation in the radio flux as a function of orbital phase
  can, in part, be traced back to changes in post-shock gas pressure,
  and thus pre-shock wind speeds
  (\S~\ref{subsec:radio_contributions}). Depending on the exact value,
  and ratio, of the energy injection parameters for relativistic
  electrons ($\zeta_{\rm rel}$) and magnetic field ($\zeta_{\rm B}$),
  the role of the different attenuation/cooling mechanisms in bringing
  about radio flux variations changes. When $\zeta_{\rm rel} \gg
  \zeta_{\rm B}$, inverse Compton cooling significantly reduces
  emission, and the Razin effect and free-free absorption contribute
  to the depth of the minimum. In contrast, when $\zeta_{\rm B} \gg
  \zeta_{\rm rel}$, the Razin effect is negligible, free-free
  absorption is the largest reducer of emission, and inverse Compton
  cooling of relativistic electrons actually {\it increases}
  emission. This latter result has not been observed in previous
  studies and relates to a subtle effect involving the characteristic
  synchrotron frequency and the shape of the spectrum through the
  function $F(\nu/\nu_c)$.
\item Using the results of the radio analysis, namely the value of the
  post-shock magnetic field strength provided by the best-fit
  $\zeta_{\rm B}$, we estimate the surface magnetic field strength to
  be $\simeq 0.3-52\;$G, which lies below the observed values for
  strongly magnetized single stars. 
\end{itemize}

We close with a note that the results of the radio models indicate a
better match to observations when values of $p \ll 2$ are
adopted. Such values deviate from the standard test particle result of
$p=2$, and hence can provide insight into the physics occurring at the
shocks and the mechanism that accelerates the non-thermal
particles. One possibility would be that the particles are accelerated
by diffusive shock acceleration and values of $p \ll2$ are the result
of shock modification by upstreaming particles. However, re-arranging
Eq~\ref{eqn:nu_char} for $\gamma$ and evaluating the terms using the
hydrodynamic simulation, and noting that $B \propto (\zeta_{\rm B}
P)^{1/2}$, we find that for representative values of $\zeta_{\rm B}=
5\times10^{-5}$ and 0.5, $\gamma$ is of the order of $10^1 -
10^2$. For electrons of this relatively low energy, shock modification
would produce values of $p>2$. Hence, the model results are {\it not }
consistent with shock modification. As such, values of $p \ll 2$ must
be produced by some alternative mechanism, e.g. re-acceleration at
multiple weak shocks - see \cite{Pittard:2006} and references therein
for further discussion of these points. Further analysis will be
required to assess the acceleration mechanism in more detail.

\begin{acknowledgements} 
  We thank the referee for a useful report that helped to improve the
  clarity of the paper. This research has made use of software which
  was in part developed by the DOE supported ASC/Alliance Center for
  Astrophysical Thermonuclear Flashes at the University of
  Chicago. Y.~N. thanks FNRS and PRODEX for funding.
\end{acknowledgements}


\end{document}